\documentclass[pre,twocolumn,floatfix,aps]{revtex4}
\usepackage{amsmath}
\usepackage{makeidx}
\usepackage{amssymb}
\usepackage{graphicx}

\usepackage[usenames]{color}
\usepackage[normalem]{ulem}

\usepackage{hyperref}
\usepackage[T1]{fontenc} 

\begin{document}

\title{Symbiotic solitons in a quasi-one- and quasi-two-dimensional spin-1 condensates}

\author{S. K. Adhikari\footnote{sk.adhikari@unesp.br      \\  https://professores.ift.unesp.br/sk.adhikari/ }}

\affiliation{Instituto de F\'{\i}sica Te\'orica, Universidade Estadual Paulista - UNESP, 01.140-070 S\~ao Paulo, S\~ao Paulo, Brazil}


\date{\today}
\begin{abstract}
We study the   formation of  spin-1 symbiotic spinor solitons 
 in a 
quasi-one- (quasi-1D) and quasi-two-dimensional (quasi-2D)  hyper-fine spin $F=1$ ferromagnetic 
Bose-Einstein 
condensate (BEC).  The symbiotic solitons necessarily have a repulsive intraspecies interaction and are bound due to an attractive interspecies interaction.
Due to a collapse instability in higher dimensions, an additional spin-orbit coupling is
necessary to stabilize a quasi-2D symbiotic spinor soliton. Although a quasi-1D symbiotic soliton  has a simple 
Gaussian-type density distribution, novel spatial periodic structure in density is found in 
quasi-2D symbiotic SO-coupled spinor solitons. 
 For a weak SO coupling, the quasi-2D solitons are 
of the $(-1, 0, +1)$ or $(+1, 0, -1)$ type with intrinsic vorticity and multi-ring structure, for Rashba or Dresselhaus SO coupling, respectively, where
the numbers in the parentheses are angular momenta projections in spin components $F_z = +1, 0, -1$,
respectively. For a strong SO coupling, stripe and superlattice solitons, respectively,
with a stripe and square-lattice modulation in  density, are found in addition to the multi-ring  solitons.
The stationary states were obtained by  imaginary-time propagation  of a mean-field model;
 dynamical stability of the solitons was established by real-time propagation over a long period of time. The possibility  of the creation of such a soliton by removing the trap of a confined  spin-1  BEC  
in a laboratory is also demonstrated.

\end{abstract}

\maketitle

\section{Introduction}
\label{Sec-I}
A bright soliton \cite{Kivshar} or a soliton  is a  self-bound solitary wave that  can move at a 
constant velocity maintaining  its shape due to a cancellation 
of   non-linear  attraction and linear repulsion. Bright solitons have been created and studied  in a   $^7$Li \cite{li} and $^{85}$Rb \cite{rb} Bose-Einstein condensate (BEC)
 by controlling  the  non-linear attraction  near a 
Feshbach resonance \cite{Inouye}, so that the atomic scattering length is changed to a desired negative value.
A vector soliton is a solitary wave with multiple components  that maintains its shape during propagation. An ordinary soliton  has effectively only one (scalar)   component or species, while vector solitons have more than one  distinct species. Most of the vector solitons studied so far are self attractive, so that, due to an intraspecies attraction, vector solitons can be realized independent of the nature of the interspecies interaction: attractive or repulsive.  However, it is also possible to create a self-repulsive binary (two-component) vector soliton 
where the intraspecies interaction is repulsive \cite{Perez-Garcia}. In that case a vector  soliton is formed due to an
interspecies attraction where the presence of different species is fundamental for the formation of the vector soliton.  Such a self-repulsive vector soliton  is usually  called a symbiotic  soliton.  

Most studies of a symbiotic soliton involve only two 
distinct species, because it is difficult to realize  a three-component hetero-nuclear BEC experimentally.
After the experimental observation of a  hyper-fine spin 
$F=1$  \cite{exp} spinor BEC of $^{23}$Na atoms,  self-attractive pseudo  spin-1/2 spinor solitons have been 
extensively studied in spinor BECs  \cite{Ieda}.
  The interspecies interaction in a spin-1 spinor BEC is very different from  that in a 
three-component hetero-nuclear BEC and it is not {\it a priori} clear that a symbiotic soliton can be formed in a spin-1 spinor BEC.   
In this paper we demonstrate that  a dynamically-stable symbiotic soliton can indeed  be formed in a 
quasi-one-dimensional (quasi-1D) spin-1 spinor BEC and study its statics and dynamics.  The viability of the formation of a quasi-1D spin-1 symbiotic soliton  by removing the trap of a confined spin-1 BEC in a laboratory is also demonstrated.

A quasi-two-dimensional (quasi-2D) or a three-dimensional  soliton cannot be stabilized in a BEC due to a collapse instability \cite{Kivshar,coll}. However, it has been demonstrated that  a quasi-2D soliton can be formed in a pseudo spin-1/2 spinor BEC in the presence of a spin-orbit (SO) 
coupling \cite{malo}. There have been studies of solitons in a quasi-1D pseudo spin-1/2 SO-coupled BEC \cite{rela}, in a quasi-1D spin-1  SO-coupled BEC \cite{sol1d}   and in a quasi-2D spin-1  SO-coupled BEC \cite{sol2d}.
Although, there could not be a natural SO coupling in an atomic  spinor BEC, 
an artificial synthetic   SO coupling can be realized in such a  BEC using   tuned Raman lasers coupling the different spin states  \cite{stringari,rev}.   Two such possible   
SO couplings are due to  Rashba \cite{Rashba} and 
Dresselhaus \cite{Dresselhaus} and other types of SO coupling are also possible.  An equal mixture of
 Rashba and Dresselhaus SO couplings has been realized experimentally in a pseudo spin-1/2 $^{87}$Rb \cite{11} and $^{23}$Na
\cite{12} BEC of $F_z = 0, -1$ spin states. Later, an SO-coupled
spin-1 $^{87}$Rb BEC of $F_z = \pm 1, 0$ states was observed and investigated \cite{13} following a suggestion \cite{SOspin1}. 

In view of these, we  explore the possibility of the formation of a quasi-2D symbiotic SO-coupled spin-1  spinor soliton  and demonstrate that such a soliton can indeed be created with  different spatial periodic patterns in density. 
For a weak SO coupling,   $(\mp  1,0, \pm 1)$-type \cite{kita}  multi-ring solitons are found for a Rashba 
or a Dresselhaus SO coupling, where the upper (lower) sign corresponds to Rashba (Dresselhaus) SO coupling.  These solitons have  angular momenta  projections $(\mp 1 , 0, \pm 1 )$
  at the center of $F_z= +1,0,-1$ components, respectively, where the positive (negative) sign inside the parenthesis denotes a vortex (antivortex).  For a larger SO coupling, two different types of quasi-degenerate solitons are found. In the first type,  the multi-ring solitons evolve into a new type of metastable solitons
with broken angular symmetry.  In the second type, the solitons develop a spatially-periodic pattern in density in the form of a stripe or a square-lattice modulation. These solitons usually have very large spatial extention.
  In all cases the soliton densities and energies are the same for both Rashba and Dresselhaus SO couplings 
although the two wave functions are different. 
 We demonstrate the possibility of the formation of a quasi-2D symbiotic
SO-coupled spin-1  spinor soliton in a laboratory by suddenly removing the trap of a trapped BEC. 

 {
A quasi-2D symbiotic
SO-coupled spin-1  spinor soliton, with a periodic stripe or a square-lattice  modulation in density,
 is quite similar to a supersolid} \cite{sprsld}. A supersolid is a quantum state, where matter forms a  periodic rigid structure, breaking continuous translational symmetry, as in a crystalline solid, and  enjoying 
friction-less flow, as in a superfluid, breaking Gauge symmetry.  Supersolidity has been suggested \cite{26} and observed \cite{27} in a dipolar BEC. In a pseudo spin-1/2 SO-coupled BEC of $^{23}$Na atoms, supersolidity was   realized experimentally  \cite{28} in the form of a quasi-1D stripe pattern
in density.  In a quasi-2D symbiotic
SO-coupled spin-1  spinor   
stripe soliton, only the component densities acquire a 2D stripe modulation, whereas in a superlattice soliton both component and total densities develop a 2D square-lattice modulation.
The present quasi-2D symbiotic spin-1 soliton with a 2D square-lattice pattern in both component and total densities, sharing properties with a conventional
supersolid, will be termed a symbiotic superlattice soliton following a previous suggestion   \cite{adhik,29}. {The quasi-2D symbiotic spin-1 soliton with a stripe pattern only in component densities will be called a stripe soliton, although such a state has often been called a super-stripe state in the literature.
 The superlattice spin-1 spinor soliton is a quasi-2D generalization  \cite{adhik} of the  quasi-1D super-stripe pseudo spin-1/2 trapped BEC \cite{29} with a stripe pattern in total density. 
 The multi-ring  solitons only exhibit
a quasi-periodic structure in the component densities without a prominent 
periodic structure in total density.

A spin-$1$ spinor BEC is controlled  by two interaction strengths, e.g.,
  $c_0\propto (a_0+2a_2)/3$ and $c_2 \propto (a_2-a_0)/3$, with $a_0$ and 
$a_2$ the scattering lengths in total spin $F = 0$ and 
2 channels,  respectively \cite{Ohmi}. All  spin-1 spinor BECs can be classified into  two distinct types \cite{Ohmi,stringari}: 
ferromagnetic    ($c_2<0$) and anti-ferromagnetic ($c_2>0$).  
 In this paper, we study  three-component spin-1 symbiotic vector solitons in quasi-1D and quasi-2D traps  using a
numerical solution of the respective mean-field  coupled Gross-Pitaevskii (GP)
equation \cite{gp} by imaginary-time simulation. These symbiotic solitons should  have   intraspecies repulsion, requiring $c_0> 0$. 
In addition, if we take  identical densities in spin components $F_z= \pm 1$ and  $c_2<0$ (ferromagnetic), 
it is possible to have a quasi-2D symbiotic spin-1 spinor  soliton  of the ferromagnetic type
bound by 
interspecies 
attraction,  although some interspecies  interactions  are repulsive.  For a quasi-1D 
symbiotic spin-1 spinor soliton, we present an approximate analytic solution in good agreement with the 
numerical result.   We study the dynamics of the symbiotic  soliton numerically by real-time simulation and establish its  dynamical stability.

  In Sec. \ref{A}  and \ref{B}, we describe the mean-field model GP equation 
for a quasi-1D symbiotic   tertiary (three-component)  BEC soliton  and a  quasi-1D symbiotic  spin-1 spinor  BEC soliton, respectively. We also provide an analytic  solution of the model for these  cases.
In Sec.  \ref{C},  we present the mean-field GP equation for a quasi-2D symbiotic SO-coupled spinor soliton.   In 
Sec. \ref{3A},   numerical result  for energy and density of the stationary  quasi-1D symbiotic   spin-1 spinor   BEC soliton is compared with the analytic approximation. In Sec. \ref{3B}, numerical result for energy and density of a  quasi-2D symbiotic  
SO-coupled  spin-1 spinor   multi-ring, stripe and superlattice 
BEC soliton   is presented. 
In Secs. \ref{3C} and \ref{3D} dynamical stability of a quasi-1D symbiotic  spin-1 spinor  BEC soliton and a quasi-2D symbiotic SO-coupled spin-1  spinor 
BEC soliton is established numerically. In Sec. \ref{3C}, we also demonstrate that a  symbiotic spin-1 spinor  BEC soliton can be created in a laboratory by removing the trap of a confined BEC with the same parameters. 
  In Sec. \ref{Sec-IV}  a summary of our findings is presented.


\section{Theoretical Formulation}
\label{Sec-II}

\subsection{Quasi-1D symbiotic tertiary soliton}
\label{A}
In a multi-component BEC, most studies on solitons are in self-attractive systems where the 
intraspecies nonlinearities are attractive. Here we consider the formation of   a symbiotic 
tertiary soliton in a self-repulsive three-component BEC, where the intraspecies nonlinearities 
are repulsive; the soliton is bound by the attractive interspecies nonlinearities.  Similar 
binding is possible in a symbiotic binary soliton \cite{Perez-Garcia}.  A tertiary BEC is described 
by the following GP equation, for the wave-function components $\psi_i(x,t),$ $i=1,2,3$, in dimensionless units: 
\begin{equation}
 \text i {\partial_t \psi_{ i}(x,t)} =
 \big[ -{\textstyle \frac{1}{2}}{\partial_x^2}
 +   {g}_0 |\psi_i|^2  + g'{\textstyle \sum}_{\bar i \ne i}|\psi_{\bar i}|^2 \big]\psi_{i}(x,t),\label{eq-1} 
\end{equation}
where $\text i=\sqrt{-1}$,
the partial space and time derivatives are denoted $\partial_x\equiv \partial /\partial x$ and $\partial_t\equiv \partial /\partial t$ and $g_0$ $(>0)$ and $g'$ $(<0)$ 
are repulsive intraspecies and attractive interspecies nonlinearities, respectively.  For simplicity we have taken the  
 intraspecies and interspecies nonlinearities 
to be equal for different components and this simplifies 
the analytic treatment but otherwise has no effect on the formation of a symbiotic tertiary soliton.   
The time-independent stationary solution $\psi_i(x)$
satisfies $\psi_i(x,t) \equiv \psi_i(x) \exp(-\text i\mu t)$ with
\begin{equation}
 \mu \psi_{ i}(x) =
 \big[\textstyle -\frac{1}{2}{d_x^2}
 +   {g}_0 |\psi_i|^2  + g'{ \sum}_{\bar i\ne {i}}|\psi_{\bar i}|^2 \big]\psi_{i}(x),\label{eq-0} 
\end{equation}
where $d_x =d/dx$ and $\mu $ is the chemical potential.
As in this case the three components are equivalent, 
 the spatial dependence of the three components will be the same:
\begin{align}\label{eq-2}
  \psi_i(x)\equiv \textstyle \frac{1}{\sqrt 3}\Phi(x), \quad i=1,2,3;
\end{align}
  so that $|\Phi(x)|^2  = \sum_i  |\psi_i(x)|^2$, where $\Phi(x)$ satisfies the following GP equation
\begin{align}
  {\mu \Phi(x)} &=
 \big[ -\textstyle\frac{1}{2}{d_x^2}
 +   {g} |\Phi|^2   \big]\Phi(x).\label{eq-3} 
\end{align}
The consistency between Eqs. (\ref{eq-0})-(\ref{eq-3}) requires $g=(g_0+2g ')/3$.  
As  the intraspecies nonlinearity $g_0$ is repulsive (positive), for a solitonic solution the interspecies nonlinearity $g '$  must be {\it attractive} (negative) and also  $2|g'|> g_0$ so that the  effective nonlinearity $g$ is attractive (negative).  
The following secant hyperbolic function is the normalized analytic solution to Eq. (\ref{eq-3}) \cite{malomed}
\begin{equation}\label{sech}
\Phi(x) =  \frac{\sqrt {|g|}}{ 2}\mbox{sech} \left(\frac{|g| x}{2}\right).
\end{equation}
The energy functional of this solution yields the following analytic estimate of energy 
 \begin{align} 
E_a=\textstyle \frac{1}{2}{\textstyle \int_{-\infty}^{\infty} }dx  \left[|d_x\Phi(x)|^2 + g |\Phi(x)|^4 \right]=-\frac{g^2}{24}.
\end{align}

\subsection{Quasi-1D symbiotic spin-1 soliton}
\label{B}

\subsubsection{Mean-field equation}

We consider a quasi-1D spin-1 spinor BEC along the $x$ axis 
realized by  strong harmonic traps in the $y-z$ plane, so that 
the system is frozen in the Gaussian ground state in these directions  \cite{Salasnich}.  
A quasi-1D  spin-1 BEC of $N$ atoms
can be described by the following set of three-coupled mean-field partial 
differential GP equations for the wave-function components $\psi_i(x,t),$ $i=+1,0,-1$ in dimensionless units 
\cite{Ohmi,sol1d}
\begin{align}
\text i {\partial_t \psi_{\pm 1}} &=
 \big[ -{\textstyle\frac{1}{2}} \partial_x^2
 +   {c}_0 {n}\big]\psi_{\pm 1} \nonumber\\ 
 &+c_2[ {n}_{\pm 1}+ {n}_0- {n}_{\mp 1}]\psi_{\pm 1}+ c_2 \psi_0^2\psi_{\mp 1}^*,\label{gps-1}\\
\text i{\partial_t \psi_0}  &= 
 \big[ -\textstyle\frac{1}{2}\partial_x^2
 + {c}_0 {n}\big]\psi_0   \nonumber\\
  &  
  +c_2[ {n}_{+1}+ {n}_{-1}]\psi_0+ 2 {c}_2\psi_0^*\psi_{+1}\psi_{-1},\label{gps-3}
\end{align}
where the interaction strengths are \cite{abc2}
\begin{align} \label{c0c2-1}
 {c}_0 = \frac{2N (a_0+2a_2)l_0} {3l^2_{yz}} , \quad 
 {c}_2 =\frac{ 2N (a_2-a_0)l_0} {3l^2_{yz}},
\end{align} 
and  component density 
$ {n}_i(x) = |\psi_i(x)|^2$ with $i=+1,0,-1$ corresponding to the three spin components $F_z=+1,0,-1$,  
 total density ${n}(x) = \sum n_{i}(x)$;  $l_{yz}$ is the harmonic oscillator length in the transverse $yz$ plane and $l_0$  is an arbitrary  scaling length in the $x$ direction.  The length is measured in units of $l_{0}$ and densities in units of $l_{0}^{-1}$. 
 A spin-1 spinor BEC is
classified into two magnetic phases: ferromagnetic
($c_2 < 0$) and antiferromagnetic ($c_2 > 0$).
The total density is  normalized to unity, i.e., 
$
 \int {n}( {x})d {x} = 1. 
$ The conserved magnetization is $m=\int 
dx  [n_{+1}(x)- n_{-1}(x)] .$  The time-independent stationary state $\psi_i(x)$ satisfies $\psi_i(x,t)\equiv \psi_i(x) \exp(-\text i\mu_it)$, with
\begin{align}
\textstyle \mu_{\pm 1} \psi_{\pm 1} &=
 \big[ -\textstyle\frac{1}{2} d_x^2
 +   {c}_0 {n}\big]\psi_{\pm 1} \nonumber\\ 
 &+c_2[ {n}_{\pm 1}+ {n}_0- {n}_{\mp 1}]\psi_{\pm 1}+ c_2 \psi_0^2\psi_{\mp 1}^*,\label{gps-10}\\
 \mu_0 \psi_0  &= 
 \big[ -\textstyle\frac{1}{2} d_x^2
 + {c}_0 {n}\big]\psi_0   \nonumber\\
  &  
  +c_2[ {n}_{+1}+ {n}_{-1}]\psi_0+ 2 {c}_2\psi_0^*\psi_{+1}\psi_{-1},\label{gps-30}
\end{align}
where $\mu_i$ is the chemical potential.
 The energy functional for a stationary state described by the mean-field GP equations 
(\ref{gps-10}) and (\ref{gps-30}) is
\cite{Ohmi}
\begin{align} \label{energy}
 E &=\textstyle \frac{1}{2} \int_{-\infty}^{\infty} dx \big[\sum_j\left|{d_x\psi_{j}}
  \right|^2   +  c_0n^2  + c_2 \big\{  n_{+1}^2+n_{-1}^2 \nonumber\\
&+  2\left(n_{+1} + n_{-1} \right)
  n_0   -2  n_{+1}n_{-1}
  \nonumber\\
   &+ 
  2\left( \psi_{-1}^*\psi_0^2\psi_{+1}^*\right. 
  \left.+\psi_{-1}\psi_{0}^{*2}\psi_{+1}\right)  \big\}
   \big].
\end{align}

\subsubsection{Analytic Consideration}

Numerical calculation for the ground-state densities of
 a ferromagnetic BEC $(c_2 < 0)$ has revealed that
the component densities are essentially multiples of each
other according to \cite{abc2}
\begin{align}\label{cond}
\psi_i(x)=\alpha_i\Phi(x), \quad  i=\pm 1,0, 
\end{align}
where $\alpha_j$ is in general complex.  Equation (\ref{cond}), when substituted in Eqs. (\ref{gps-10})-(\ref{gps-30}), 
leads to three equations for the same function $\Phi$. A consistency between these 
three equations  require that  $\Phi(x)$ satisfies  Eq. (\ref{eq-3}) with the analytic solution
(\ref{sech})
with $g=c_0+c_2$ and  \cite{abc2}
\begin{equation}
\psi\equiv  \left( \begin{array}{c} \psi_{+1}\\
\psi_0\\
\psi_{-1}
 \end{array} \right)
= \frac{1}{2}\left( \begin{array}{c}
(1+m)\Phi(x) \\
\sqrt{2(1-m^2)} \Phi(x)\\
(1-m)\Phi(x) \end{array} \right).
\label{Gaussian}
\end{equation} 
For a self-repulsive quasi-1D  symbiotic spin-1 BEC soliton, $c_0$ must be positive (repulsive) and 
$c_2$ has to be negative (attractive), in addition to the condition $g\equiv c_0+c_2<0$ necessary for the system 
to have overall attraction. 
But, for a negative $c_2$, in Eq. (\ref{gps-1}) there is the self-attractive term
$c_2n_{\pm 1} \psi _{\pm 1}$  in components $i=\pm 1$.  However, if we are limited to the solution with  the property $n_{+1}= n_{-1}$, 
leading to zero magnetization ($m=0$),    Eq. (\ref{gps-1}) becomes    
\begin{eqnarray}
\textstyle\text  i {\partial_t \psi_{\pm 1}} =
  \big[ -\frac{1}{2}{\partial_x^2}
 +   {c}_0 {n}\big]\psi_{\pm 1}
 +c_2 {n}_0\psi_{\pm 1}+ c_2 \psi_0^2\psi_{\mp 1}^*.\label{gps-2} 
\end{eqnarray}
Hence, all solutions of Eqs. (\ref{gps-3}) and (\ref{gps-2}) for $c_0>0$ and $c_2 <0$
 satisfying  $n_{+1}(x) = n_{-1}(x)$
represent quasi-1D symbiotic spin-1 spinor solitons, where a positive $c_0$ and negative $c_2$ correspond to intraspecies repulsion and interspecies attraction.  Some interspecies interactions (proportional to $c_0$) in this model are repulsive. Nevertheless, all interspecies interactions 
(proportional to $c_2$) are  attractive and the symbiotic vector soliton is formed due to these interspecies attractions. As $\Phi(x)$ satisfies Eq. (\ref{eq-3}) 
with $g=c_0+c_2$, analytic approximation (\ref{Gaussian})  for  $n_{+1}(x) = n_{-1}(x)$ and $m=0$, appropriate for a quasi-1D symbiotic spin-1 spinor soliton, becomes 
 \begin{eqnarray}
\left( \begin{array}{c} \psi_{+1}(x)\\
\psi_0(x)\\
\psi_{-1}(x)
 \end{array} \right)
= \frac{\sqrt{|c_0+c_2|}}{4}\mbox{sech}\left[ \frac{|c_0+c_2|x}{2}\right]  \left( \begin{array}{c}
1 \\
\sqrt{2} \\
1 \end{array} \right),
\label{1dapp}
\end{eqnarray} 
with the analytic energy
\begin{equation}
\label{e_a}
 E_a=-\textstyle\frac{1}{24}(c_0+c_2)^2.
 \end{equation}

\subsection{Quasi-2D symbiotic SO-coupled spin-1 soliton}
\label{C}

\subsubsection{Mean-Field Equation}

Because of  a collapse instability, a quasi-2D soliton cannot be stabilized \cite{coll}.  However, a stabilized quasi-2D symbiotic spinor soliton is possible in the presence of an SO coupling.    
We consider a quasi-2D SO-coupled spin-1 spinor BEC in the $x-y$ plane realized by a strong harmonic trap along the $z$ direction, so that the system is frozen in the Gaussian ground state along $z$ axis. The single-particle Hamiltonian of the SO-coupled BEC is \cite{11}
\begin{align}\label{sph2}
H_0^{2D}=-\textstyle\frac{1}{2}\nabla^2_{\bf r}-i \gamma [\eta \partial_y\Sigma_x -\partial_x\Sigma_y]
\end{align}
where ${\bf r}\equiv \{x,y\}, \nabla_{\bf r}=\partial_x^ 2+\partial_y ^2, $
$\gamma $ is the strength of 
Rashba or Dresselhaus SO coupling, $\eta =+1$ ($-1$) for Rashba (Dresselhaus) coupling,
$\Sigma_x$ and $\Sigma_y$ are irreducible representation of the spin-1 matrix and are given by 
\begin{eqnarray}
\Sigma_x=\frac{1}{\sqrt 2} \begin{pmatrix}
0 & 1 & 0 \\
1 & 0  & 1\\
0 & 1 & 0
\end{pmatrix}, \quad  \Sigma_y=\frac{i}{\sqrt 2 } \begin{pmatrix}
0 & -1 & 0 \\
1 & 0  & -1\\
0 & 1 & 0
\end{pmatrix}.
\end{eqnarray}  
  A quasi-2D \cite{Salasnich} SO-coupled spin-1 BEC of $N$ atoms
can be described by the following set of dimensionless three-coupled mean-field partial 
differential GP equation for the wave-function components $\psi_i({\bf r},t), i=+1,0,-1$ 
\cite{Ohmi}
\begin{align}
 \text i {\partial_t \psi_{\pm 1}}&=
 \big[ \textstyle-\frac{1}{2}\nabla_{\bf r}^2
 +   {c}_0 {n}\big]\psi_{\pm 1}  -\text i\widetilde \gamma\big[\eta \partial_y\pm \text i\partial_x\big] \psi_0\nonumber\\ 
 &+c_2[ {n}_{\pm 1}+ {n}_0- {n}_{\mp 1}]\psi_{\pm 1}+ c_2 \psi_0^2\psi_{\mp 1}^*,\label{gps-4}\\
\text i{\partial_t \psi_0} &= 
 \big[ \textstyle -\frac{1}{2}\nabla_{\bf r}^2
 + {c}_0 {n}\big]\psi_0  -\text i{ \widetilde \gamma}[-\text i\partial_x\phi_{-1}+\eta \partial_y\phi_{+1} ]
\nonumber\\
  &  
  +c_2[ {n}_{+1} + {n}_{-1}]\psi_0+ 2 {c}_2\psi_0^*\psi_{+1}\psi_{-1}  , \label{gps-5}
\end{align}
where  $\widetilde \gamma =\gamma /\sqrt 2$,
 $\phi_{\pm 1}=\psi_{+1}\pm\psi_{-1},$ and 
interaction strengths \cite{abc2}
\begin{eqnarray}
 {c}_0 = \frac{2N \sqrt{2\pi}(a_0+2a_2)} {3l_{z}} , \quad 
 {c}_2 =\frac{ 2N\sqrt{2\pi} (a_2-a_0)} {3l_{z}}, \label{c0c2-2}
\end{eqnarray}
where $l_z$ is the harmonic oscillator length in the $z$ direction.
Here all lengths are expressed in units of $l_z$ and  density in units of $l_z^{-2}$. The total density is  normalized to unity, i.e., 
$
 \int {n}( {\bf r})dx dy = 1. 
$ The  magnetization  $m=\int 
dx dy  [n_{+1}( {\bf r})- n_{-1}( {\bf r})] $ is not conserved in this SO-coupled BEC. The time-independent
version of Eqs. (\ref{gps-4})-(\ref{gps-5}), appropriate for stationary solutions, can be derived from the energy functional
\begin{align}
\label{energy2d}
E[\psi ] &= \textstyle\frac{1}{2} {\textstyle \int } dx dy \big [ {
\textstyle \sum }_{j} |\nabla _{\bf r}\psi _{j}|^{2}
+c_{0}n^{2}+ c_{2} \{n_{+1}^{2} +n_{-1}^{2}
\nonumber \\
&  +2(n_{+1}+n_{-1})n_{0}-2n_{+1}n_{-1}+2(\psi _{-1}^{*}\psi _{0}^{2}\psi _{+1}^{*}
\nonumber
\\
& + \psi _{-1}\psi _{0}^{*2}
\psi _{+1})  \}-\text i\sqrt 2{\gamma } \{ \eta \psi _{0}^{*}
{\partial} _{y} \phi _{+1}
\nonumber \\
&
+ \eta \phi _{+1}^{*} {\partial} _{y} \psi _{0} -\text i \psi _{0}^{*}
{\partial} _{x} \phi _{-1} +\text i \phi _{-1}^{*}{\partial} _{x} \psi _{0}
 \} \big ] .
\end{align}
In this case we will look for solution satisfying $n_{+1}(x)=n_{-1}(x)$ leading to  magnetization $m=0$. For this solution Eq. (\ref{gps-4}) reduces to 
\begin{align}
\text  i {\partial_t \psi_{\pm 1}}&=
 \big[ -\textstyle \frac{1}{2}\nabla_{\bf r}^2
 +   {c}_0 {n}\big]\psi_{\pm 1}  -\text i{\widetilde \gamma}[\eta \partial_y\pm\text  i\partial_x] \psi_0\nonumber\\ 
 &+c_2  {n}_0\psi_{\pm 1}+ c_2 \psi_0^2\psi_{\mp 1}^*.\label{gps-6}
\end{align}
The  symbiotic SO-coupled spin-1 soliton should satisfy Eqs. (\ref{gps-5}) and (\ref{gps-6}). If we take a positive (repulsive) $c_0$  and negative (attractive, ferromagnetic) $c_2$, all intraspecies interactions will be repulsive and the soliton will be a symbiotic one 
bound by attractive interspecies interactions. We note that in this model some interspecies interactions proportional to $c_0$ are also  repulsive.

As magnetization is not conserved in the presence of Rashba or Dresselhaus couplings, we cannot fix magnetization in the calculation and it is not clear that a solution satisfying 
 $n_{+1}(x)=n_{-1}(x)$ with $m=0$ exists.  In our numerical solution of Eqs. (\ref{gps-5}) and (\ref{gps-6}) employing imaginary-time propagation,  { we allow magnetization $m$ to freely evolve in time starting with an arbitrary  initial value $m_{\mathrm{init}}$ corresponding to the initial state used in numerical simulation. The final converged solution, so obtained, is found to have    $n_{+1}(x)=n_{-1}(x)$  and $m=0$ for all solutions reported in this paper, independent of the initial value $m_{\mathrm{init}}$, and   will  correspond to a quasi-2D symbiotic SO-coupled spinor soliton. This guarantees that the final solution   indeed satisfies Eqs. (\ref{gps-5}) and (\ref{gps-6}). 
}

{
\subsubsection{Analytic Consideration}

Many properties of an SO-coupled symbiotic spin-1 BEC can be inferred from a  study of the  eigenfunctions of the single-particle Hamiltonian in the absence of nonlinear spinor interaction. From a minimization of the SO-coupling energy, the appearance of the $(\mp 1,0,\pm 1)$-type state was demonstrated in Ref. \cite{kita}. Here we demonstrate the appearance of spatially-periodic states from a study of the single-particle Hamiltonian. 
The eigenfunction of the quasi-2D single-particle Hamiltonian (\ref{sph2})
 satisfies the following eigenvalue equation 
\begin{eqnarray}
\begin{bmatrix}\label{spf2}
-\textstyle\frac{1}{2}\nabla^2_{\bf r}-e& \widetilde \gamma \partial_{xy}^{(-)}&  0\\
-\widetilde \gamma\partial_{xy}^{(+)} &-\textstyle\frac{1}{2}\nabla^2_{\bf r}-e&\widetilde  \gamma \partial_{xy}^{(-)} \\
0& -\widetilde \gamma \partial_{xy}^{(+)} &  -\textstyle\frac{1}{2}\nabla^2_{\bf r}-e
\end{bmatrix}   \begin{pmatrix} \psi_{+1} \\ \psi_0 \\ \psi_{-1} \end{pmatrix}
=  0 ,  
\end{eqnarray}
where $\partial_{xy}^{(\pm)}=(\partial_x\pm \text i\eta \partial_y) $. Tridiagonal Eq.  (\ref{spf2}) has the following  two analytic degenerate solutions with energy  $e=-\gamma^2/2$:
\begin{align}
\label{spf3}
 \begin{pmatrix} \psi_{+1} \\ \psi_0 \\ \psi_{-1} \end{pmatrix}=\textstyle {\frac{ 1}{\sqrt 2} }  \begin{pmatrix} \cos(\gamma x) \\ -\sqrt 2 \sin(\gamma x)
 \\- \cos(\gamma x) \end{pmatrix}, \, \, 
\end{align}
and 
\begin{align}
\label{spf5}
 \begin{pmatrix} \psi_{+1} \\ \psi_0 \\ \psi_{-1} \end{pmatrix} =\textstyle {\frac{ 1}{\sqrt 2} }  \begin{pmatrix} \cos(\gamma y) \\ -\text i \eta \sqrt 2\sin(\gamma y)
 \\ \cos(\gamma y) \end{pmatrix}\, .
\end{align}
The states  (\ref{spf3})  and  (\ref{spf5})   represent stripes in density along $y$ and $x$
directions in the components with a uniform density in the
sum of the components: $|\psi_{+1}|^2+ |\psi_{0}|^2+|\psi_{-1}|^2=1$.
A linear combination of these states, e.g., 
\begin{align}
\label{spf4}
 \begin{pmatrix} \psi_{+1} \\ \psi_0 \\ \psi_{-1} \end{pmatrix} =\sqrt n   \begin{pmatrix} \cos(\gamma x) \pm  \text i \cos(\gamma y)\\ -\sqrt 2 [\sin(\gamma x)  \mp  \sin(\gamma y)]
 \\ -\cos(\gamma x) \pm \text i \cos(\gamma y)
 \end{pmatrix}, 
\end{align}
has a square-lattice pattern in component and total densities.   Similar spatially-periodic density modulations, as in Eqs. (\ref{spf3})-(\ref{spf4}), will be found in an SO-coupled symbiotic spin-1 soliton as we will see in the following.

\begin{figure}[!t]
\begin{center} 
\includegraphics[trim = 0mm 0mm 0cm 0mm, clip,width=.32\linewidth,clip]{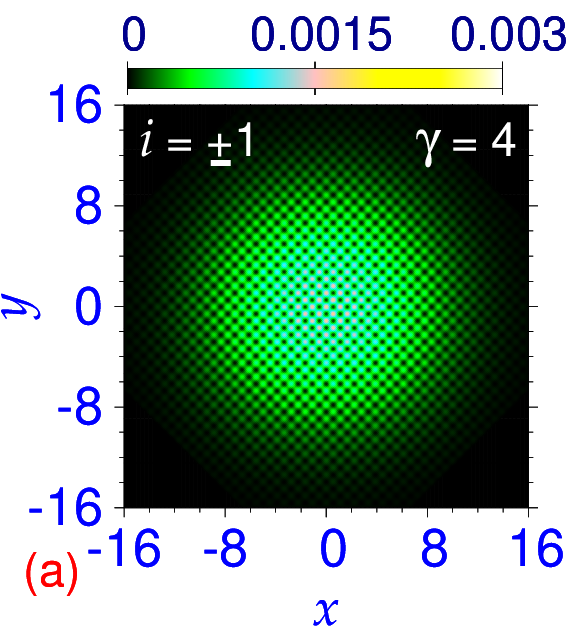}
\includegraphics[trim = 0mm 0mm 0cm 0mm, clip,width=.32\linewidth,clip]{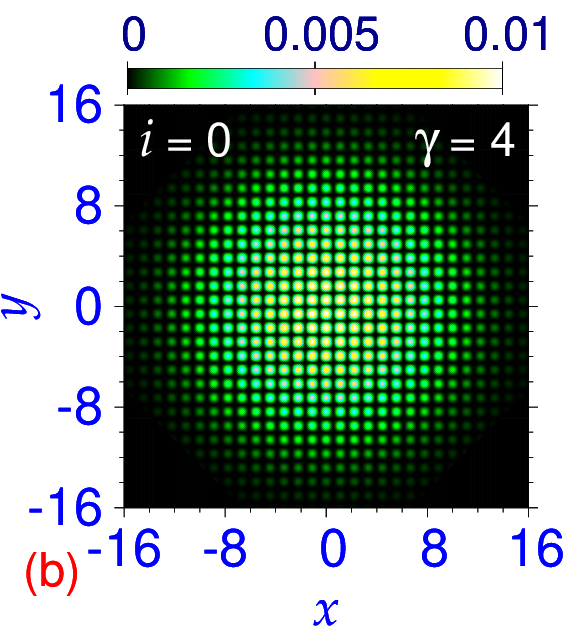}
\includegraphics[trim = 0mm 0mm 0cm 0mm, clip,width=.32\linewidth,clip]{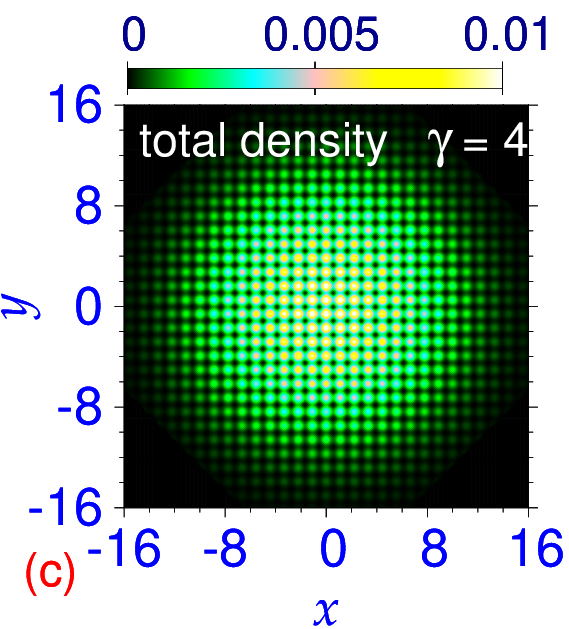}

\caption{(Color online) Contour plot of density $n_i$ of the state (\ref{spf4}) for
$\gamma  = 4$ and $\eta = +1$ of components (a) $i =\pm 1$, (b) $i = 0$,
and (c) total density after replacing the factor $\sqrt n$
by an appropriate Gaussian distribution. The densities are
normalized as 
$\int dxdy(n_{+1} + n_0+n_{-1}) = 1$. Results in all figures are
plotted in dimensionless units.}
\label{fig1} \end{center}
\end{figure}

The SO-coupled symbiotic solitons studied in this paper are localized states, whereas the solutions (\ref{spf3})-(\ref{spf4}) of the single-particle Hamiltonian  are not localized.  However, if we multiply these by an appropriate  localized 2D Gaussian state we can generate states quite similar to those of  localized stripe and superlattice solitons. This is illustrated in Fig. \ref{fig1} through a contour plot of density of the state (\ref{spf4})
 for $\gamma=4$ and $\eta =1$  of components (a) $i=\pm 1$, (b) $i=0$ and (c) total density after  replacing the factor $\sqrt n$ 
by an appropriate Gaussian.
The density of this state is quite similar to a
superlattice soliton with square-lattice pattern in density
for $\gamma =4, c_0 =1,  c_2= -2,$ viz. Figs. \ref{fig6}(a)-(c). Even the
total density in Fig. \ref{fig1}(c) has a square-lattice pattern  as in a superlattice BEC. Hence an analytic consideration of the single-particle Hamiltonian  reveals
that it can naturally lead to eigenstates with densities quite similar to 
an actual stripe  and superlattice soliton of an SO-coupled
spin-1 BEC. In a weakly-attractive uniform system,
that we consider in this paper, the energies of these states
are very close to the analytic energy $e=-\gamma^2/2$ of the states (\ref{spf3}) and (\ref{spf4})
in most
cases, indicating a negligible contribution from the nonlinear terms $c_0$ and $c_2$.

}

\section{Result and Discussion}
\label{Sec-III}

We numerically solve the coupled partial differential GP  equations  using the
split-time-step Crank-Nicolson method \cite{Muruganandam} with real- and imaginary-time propagation.
For a numerical simulation there are the FORTRAN \cite{Muruganandam} and C  \cite{cc} programs for the solution of the GP equation  and their 
open-multiprocessing \cite{omp} version  appropriate for SO-coupled spin-1 spinor BEC.
The real-time propagation method was used to study the dynamics with the converged solution     
obtained in imaginary-time propagation as the initial state. 
The space step employed for the solution of Eqs.  (\ref{gps-3}) and  (\ref{gps-2})  to obtain the   quasi-1D symbiotic solitons  by the imaginary- and real-time propagation is $dx =0.0125$ and that for the solution of 
 Eqs. (\ref{gps-5}) and (\ref{gps-6}) to obtain the   quasi-2D symbiotic SO-coupled spinor soliton  by the imaginary- and real-time propagation is $dx=dy =0.05$.  In both quasi-1D and quasi-2D cases the time step for imaginary-time propagation is $dt=0.1dx^2$  and for real-time propagation is $dt=0.05dx^2$.

{Instead of presenting results only in dimensionless units, we will also  relate the results with the most commonly used ferromagnetic $^{87}$Rb atom with the scattering lengths \cite{a0a2} $a_0=101.8a_B$ and $a_2=100.4a_B$  satisfying $c_0>0$ and $c_2<0$, where $a_B$ is the Bohr radius. In all experiments on solitons \cite{li,rb} 
the actual value of scattering length was modified by the Feshbach-resonance technique \cite{Inouye} so as to have a reasonable number of atoms in the soliton in the harmonic trap used in a laboratory. Similarly, we will  consider a   modified value of $a_2$ of  $^{87}$Rb atoms, obtained  by the  Feshbach-resonance technique in this paper:  $a_2=-20.36a_B$, appropriate for both quasi-1D and quasi-2D cases, as we will see in the following.
}

\begin{figure}[!t]
\begin{center}
\includegraphics[trim = 0mm 0mm 0cm 0mm, clip,height=4.6cm,width= 8.5cm,clip]{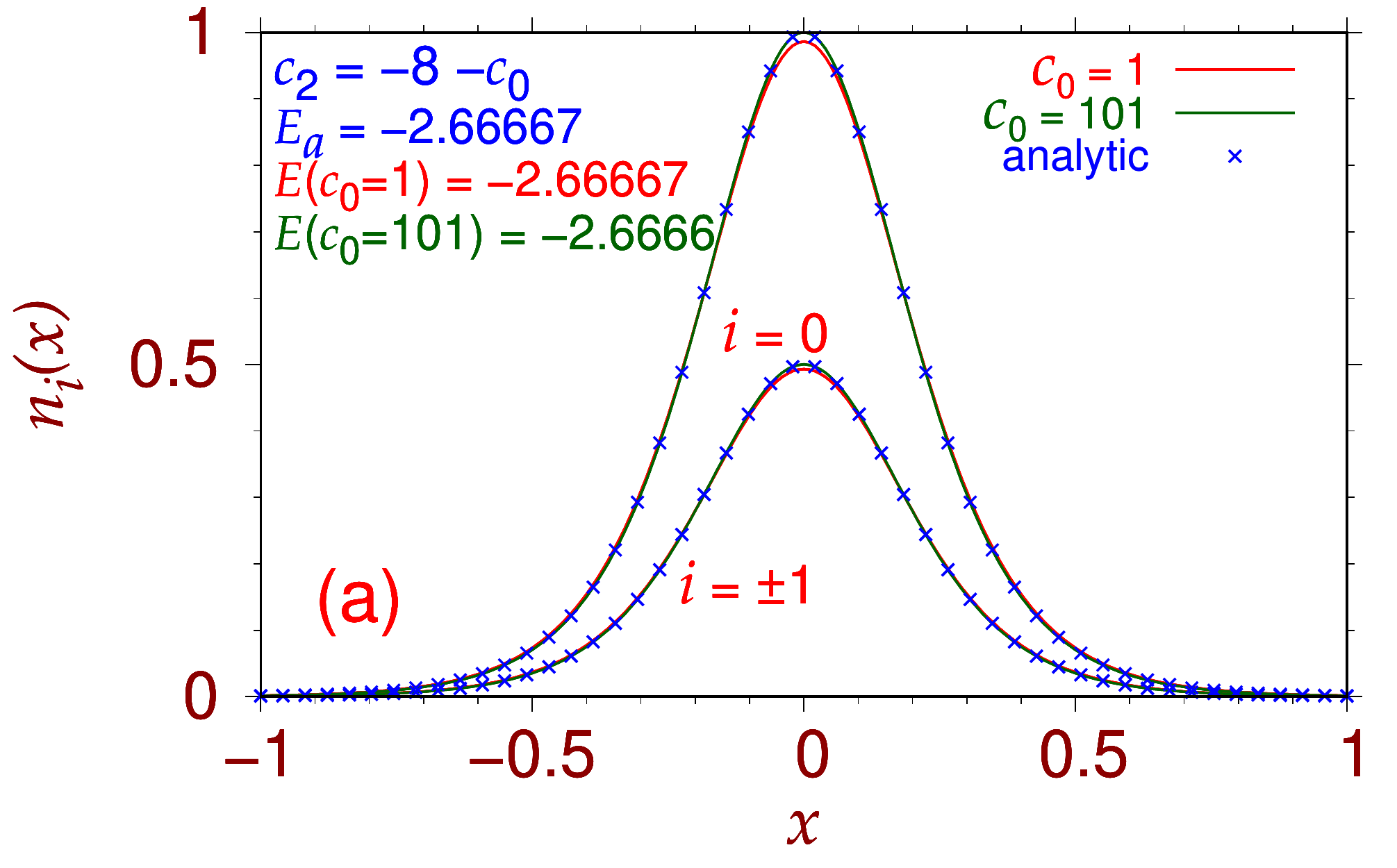}
\includegraphics[trim = 0mm 0mm 0cm 0mm, clip,height=4.6cm,width= 8.5cm,clip]{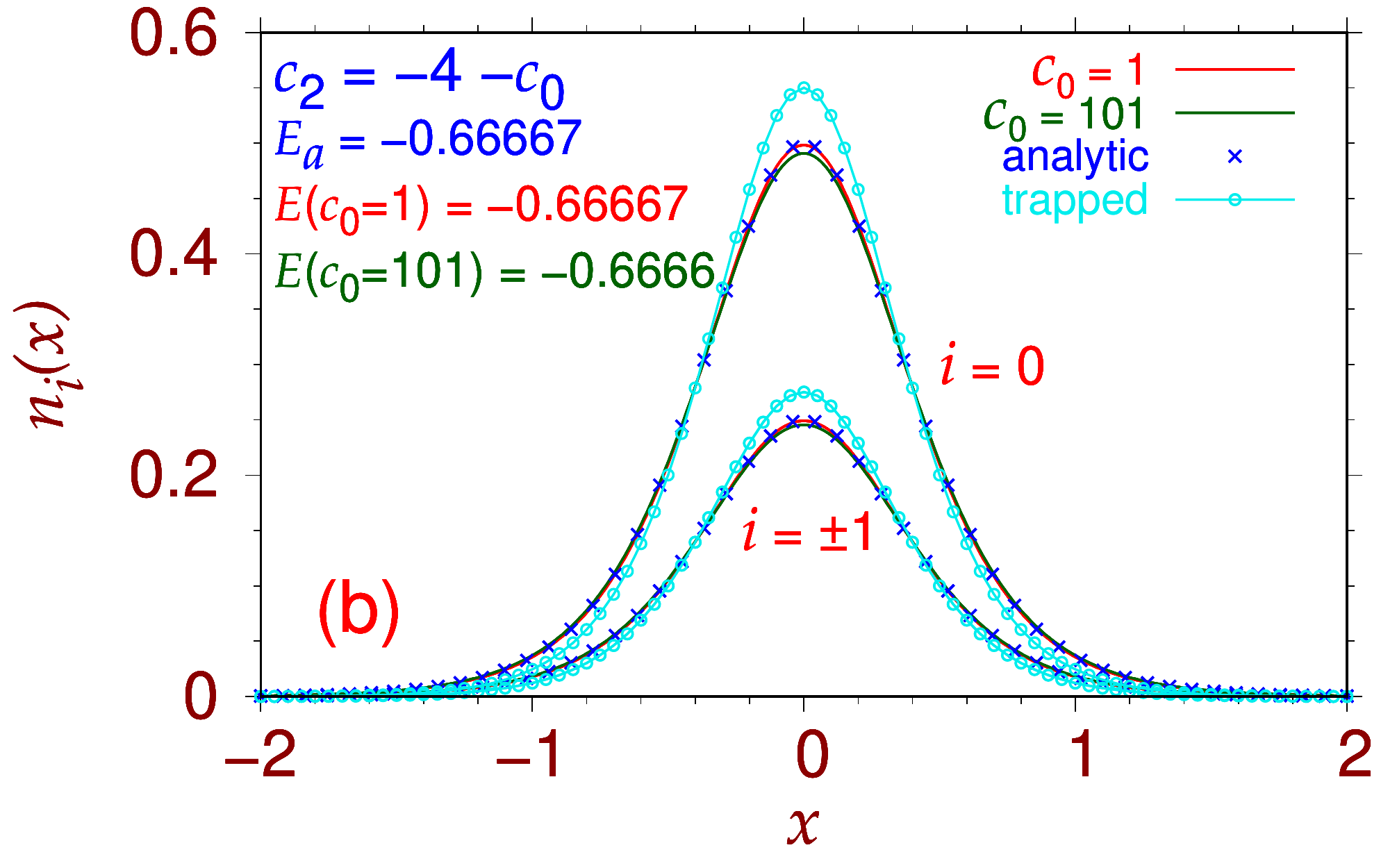}
\includegraphics[trim = 0mm 0mm 0cm 0mm, clip,height=4.6cm,width= 8.5cm,clip]{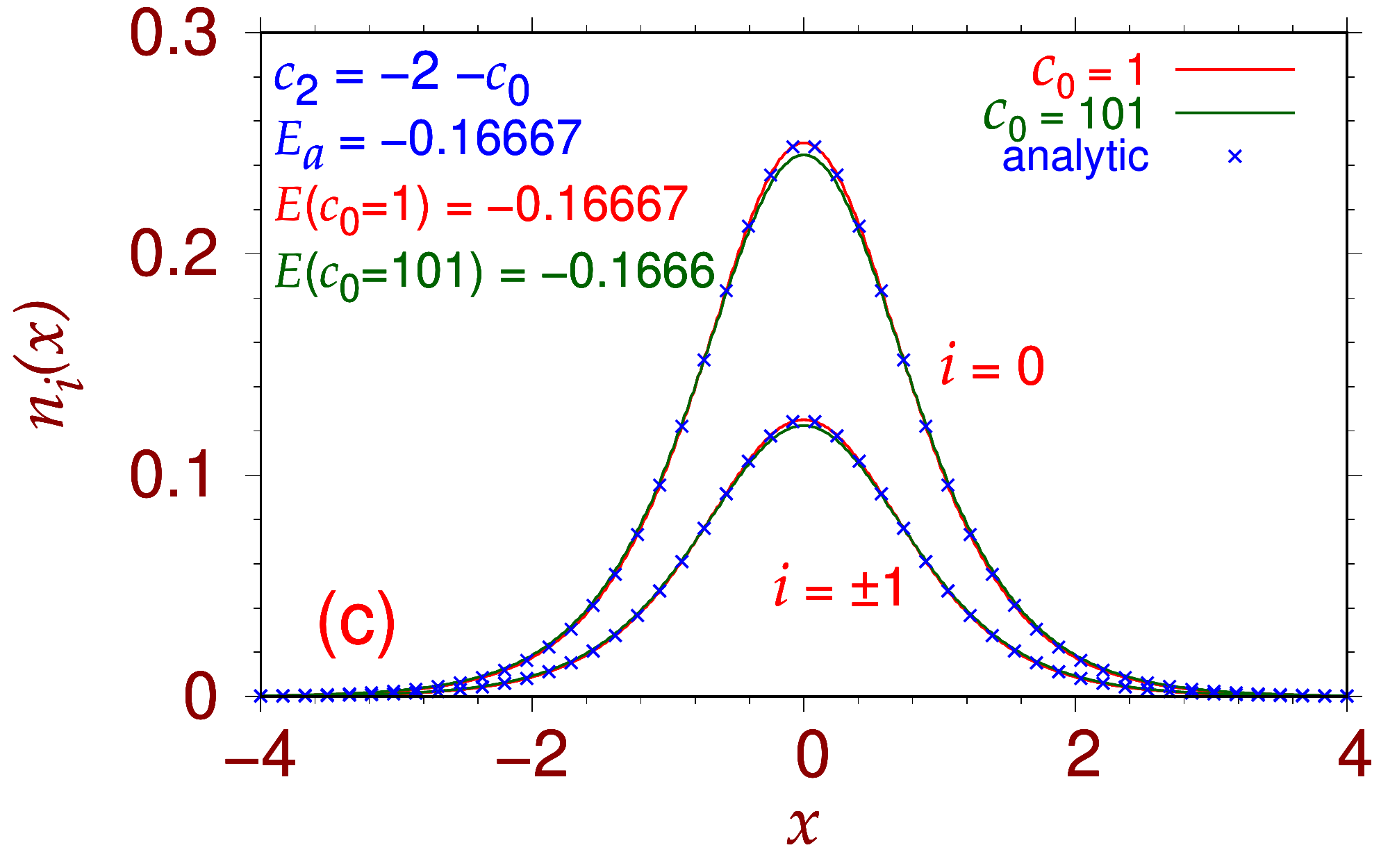}

\caption{(Color online)   Numerical (lines) and analytic (points)  densities 
$n_i(x), i =0, \pm 1,$ 
of the three components for the lowest-energy symbiotic  vector  soliton 
 with $c_0=1$ and { 101} and   (a) $c_0+c_2=-8$, (b) $c_0+c_2=-4$,  and  (c) $c_0+c_2=-2$.   
  Numerical energy $E$  and analytic energy  $E_a$ of Eq. (\ref{e_a}) are also shown. 
All quantities in this and following figures  are dimensionless.}
\label{fig2} \end{center}
\end{figure}

\subsection{Stationary soliton}

\subsubsection{Quasi-1D symbiotic spin-1 spinor soliton}

\label{3A}

For the formation of a  quasi-1D symbiotic spin-1 spinor soliton we  consider  $c_0>0$  and $c_2<0,$  such that  $c_0+c_2< 0$. In this case, magnetization is conserved (in absence  of SO coupling), hence,  in the imaginary-time propagation, we consider an initial  state with the property 
$n_{+1}(x)=n_{-1}(x)$ $(m=0)$, so that the converged final state also has the same property as required of a quasi-1D  symbiotic spin-1 spinor soliton.
The soliton profile is found to be reasonably insensitive to a variation  of $c_0$ provided that $c_0+c_2$ is kept fixed, in agreement with the analytic result (\ref{1dapp}). We demonstrate this considering $c_0=1$ and 101 corresponding to self repulsion. In Fig. \ref{fig2}(a) we illustrate the component densities of the symbiotic soliton  for $c_0+c_2=-8$  and $c_0=1$ and 101.  The analytic  density (point) and numerical density (line)
as  well as the corresponding energies are shown for both $c_0=1$ and 101.  The analytic energy $E_a$ of Eq. (\ref{e_a})  is independent of $c_0$, provided $|c_0+c_2|$ is unchanged,  and is in good agreement with the numerical energy $E$. { The  numerical energies of the quasi-1D symbiotic solitons  in Fig. \ref{fig2} are converged results without numerical error.}
 In Figs.  \ref{fig2}(b) and (c) we display the densities  for 
 $c_0+c_2=-4$  and $-2$, respectively, and $c_0=1$ and 101.  
As the value of $|c_0+c_2|$  decreases from 8 to 2,  the net attraction is reduced, consequently, 
the soliton occupies a larger extension of space as illustrated in Fig. \ref{fig2}.  In all cases, as $c_0$ increases, the numerical energy increases for a fixed   $|c_0+c_2|$.  In all cases, as required, $n_{+1}(x)=n_{-1}(x)$. 

 Let us now see if the symbiotic solitons displayed in Fig. \ref{fig2} are feasible in an actual experiment with $^{87}$Rb atoms with Feshbach-modified scattering lengths $a_0=101.8a_B$ and $a_2=-20.36a_B$.  Equation (\ref{c0c2-1}) relates the nonlinearities with the number  of atoms, scattering lengths and trap parameter.  If we take in Eq. (\ref{c0c2-1}) the parameters $l_{yz}= l_0 =2$ $\mu$m and  the  number of atoms $N=3710$ then we obtain $c_0 = 4$ and  $c_2=-8=-4-c_0$; these nonlinearity parameters with  $101>c_0>1$
correspond to the scenario of Fig. \ref{fig2}(b).   The parameter  $l_{yz}\equiv \sqrt{\hbar/M\omega_{yz}}=2 $ $\mu$m, with $M$ the mass of a $^{87}$Rb atom, yields  the average harmonic trap frequency in the $y-z$ plane:  $\omega_{yz}\approx 2\pi \times 29$ Hz.

\subsubsection{Quasi-2D symbiotic SO-coupled spin-1  soliton}

\label{3B}

Next we consider the formation of a quasi-2D symbiotic SO-coupled spin-1 spinor soliton. 
For a small $\gamma$, these solitons are of the $(\mp 1,0,\pm 1)$ type \cite{kita} where the upper (lower)
sign corresponds to Rashba (Dresselhaus) SO coupling. { In quasi-2D setting,
 Eq. (\ref{c0c2-2}) relates the nonlinearities with the number  of atoms, scattering lengths and trap parameter.  If in Eq. (\ref{c0c2-2}) we take the trap parameter $l_{z}= \sqrt{2\pi}$ $\mu$m, number of atoms $N=465,$ then we obtain $c_0 = 1$ and  $c_2=-2$; we will take these nonlinearity parameters in the study of quasi-2D symbiotic SO-coupled soliton
in this paper. In case of $^{87}$Rb atoms with Feshbach-modified scattering lengths $a_0=101.8a_B$ and $a_2=-20.36a_B$,  
the parameter $l_z\equiv \sqrt{\hbar/M\omega_z}=  \sqrt{2\pi}$ $\mu$m  leads to the following harmonic trap frequency in the $z$ direction: $\omega _z\approx 2\pi \times 18.5$ Hz.}

\begin{figure}[!t]
\begin{center}

\includegraphics[trim = 0mm 0mm 0cm 0mm, clip,width=.41\linewidth,clip]{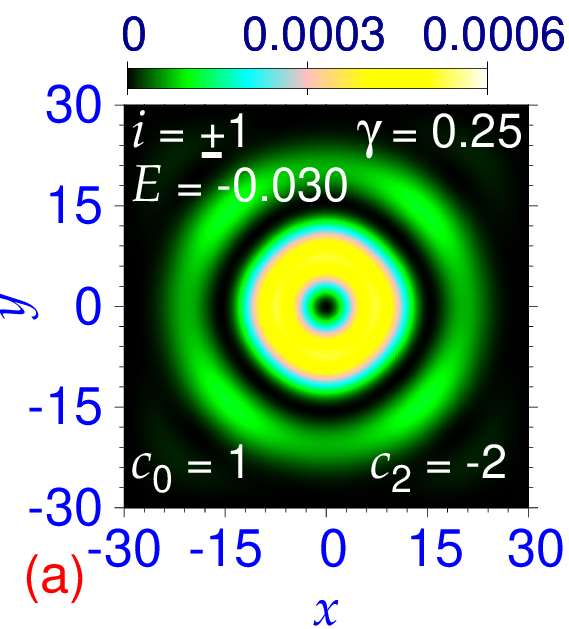}
\includegraphics[trim = 0mm 0mm 0cm 0mm, clip,width=.41\linewidth,clip]{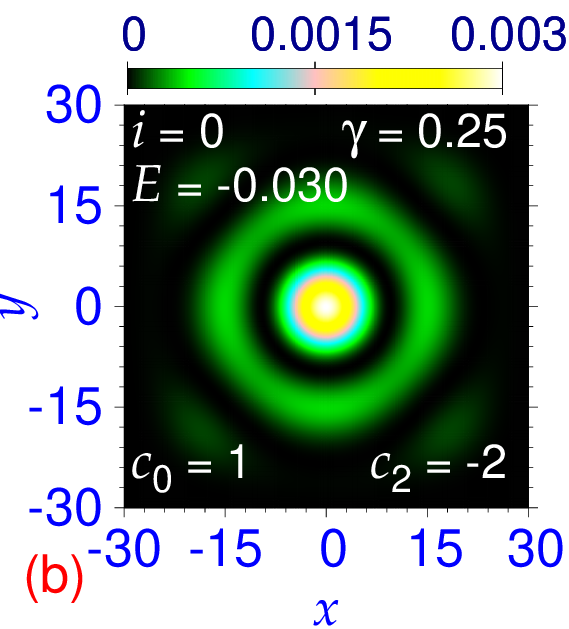}
\includegraphics[trim = 0mm 0mm 0cm 0mm, clip,width=.41\linewidth,clip]{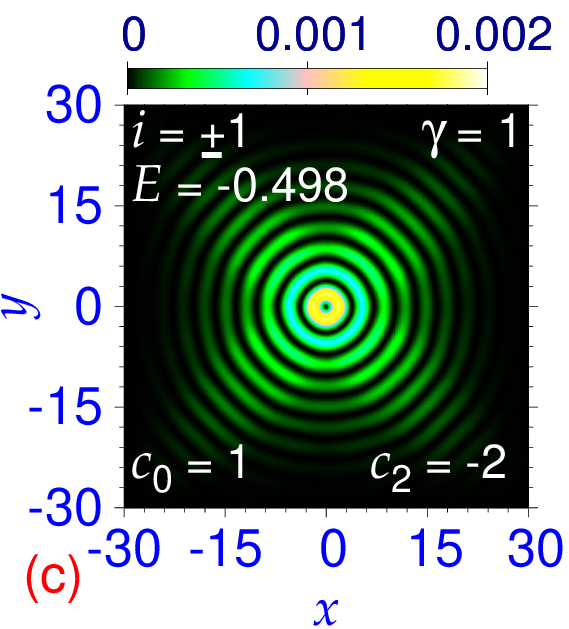}
\includegraphics[trim = 0mm 0mm 0cm 0mm, clip,width=.41\linewidth,clip]{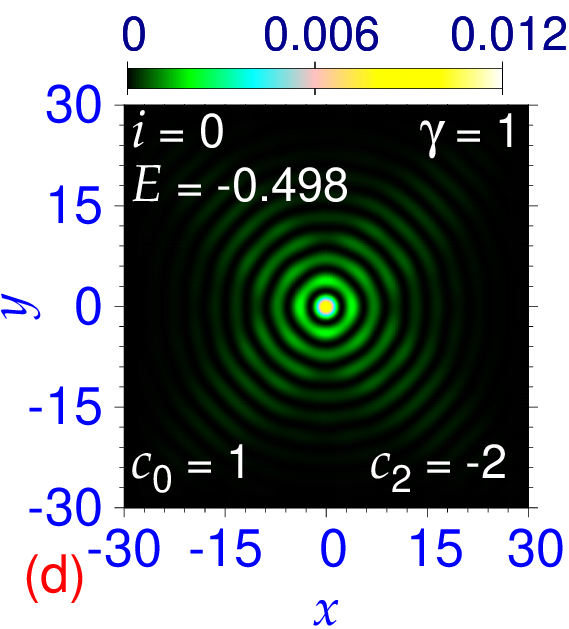}
\includegraphics[trim = 0mm 0mm 0cm 0mm, clip,width=.32\linewidth,clip]{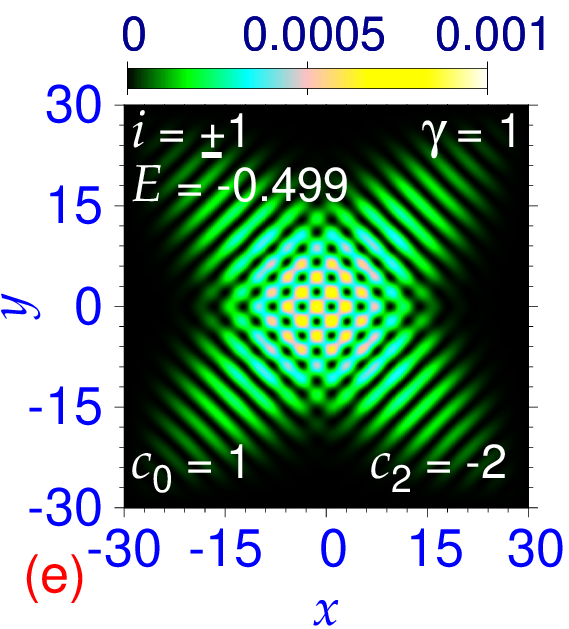}
\includegraphics[trim = 0mm 0mm 0cm 0mm, clip,width=.32\linewidth,clip]{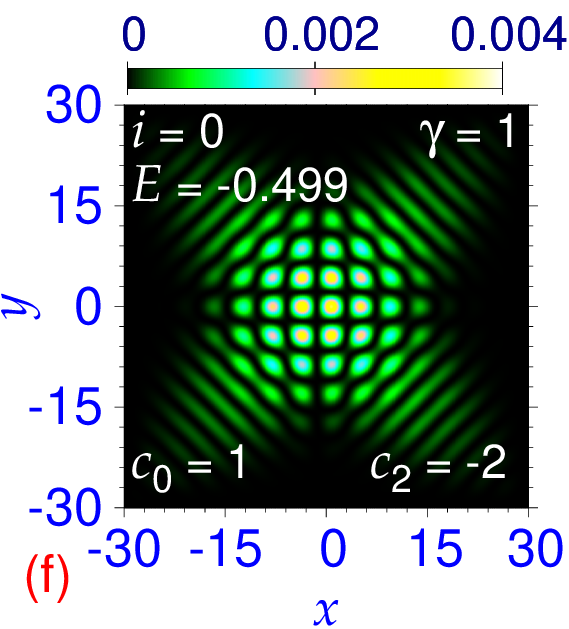}
\includegraphics[trim = 0mm 0mm 0cm 0mm, clip,width=.32\linewidth,clip]{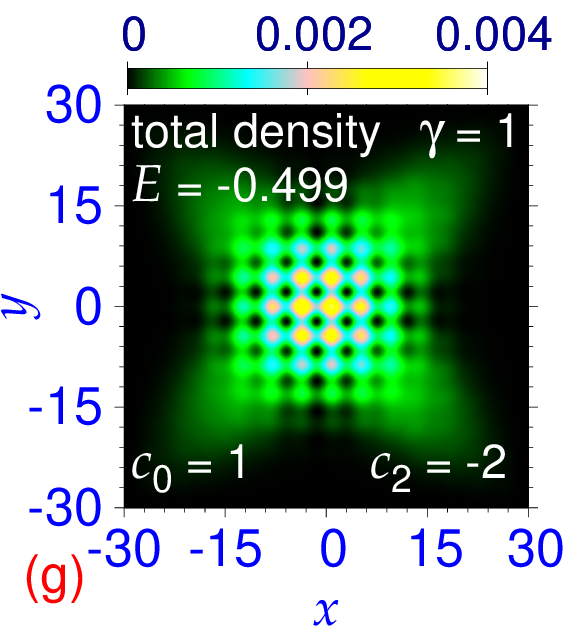}

\caption{(Color online) Contour plot of density $n_i$ of a $(\mp 1, 0, \pm 1)$-type quasi-2D symbiotic 
 Rashba or Dresselhaus SO-coupled spin-1  spinor  soliton of
components (a) $i = \pm 1$, (b) $i = 0$ with parameters $c_0=1,  c_2=-2, \gamma =0.25$ and      of a multi-ring soliton  of  components   (c) $i = \pm 1$, (d) $i = 0$  with parameters $c_0=1,  c_2=-2, \gamma =1$.   The  same  of a superlattice 
soliton with parameters  $c_0=1,  c_2=-2, \gamma =1$  of components    (e) $i = \pm 1$, (f) $i = 0$ and (g) total density.
}
\label{fig3} \end{center}
\end{figure}

 In Fig.  \ref{fig3} we display the contour plot
of density of components (a) $i = \pm 1$  and (b) $i = 0$  of a $(\mp 1, 0, \pm 1)$-type soliton for parameters
$c_0=1, c_2=-2, \gamma =0.25$. There is one prominent  ring   in this case. 
 The numerical energy ($E = -0.030\approx -\gamma^2/2=-0.03125$) and the density 
of the
$(\mp 1, 0, \pm 1)$-type soliton in Figs. \ref{fig3}(a)-(b)  are  independent of the type of SO coupling: Rashba or Dresselhaus. { The estimated numerical error in energy for all quasi-2D solitons reported in this paper is small  ($\sim \pm 0.001$).}
 In the numerical calculation by imaginary-time propagation, the vortex-antivortex structure was imprinted in the respective initial wave function components.

\begin{figure}[!t]
\begin{center}
\includegraphics[trim = 0mm 0mm 0cm 0mm, clip,height=3.cm,width= 3.cm,clip]{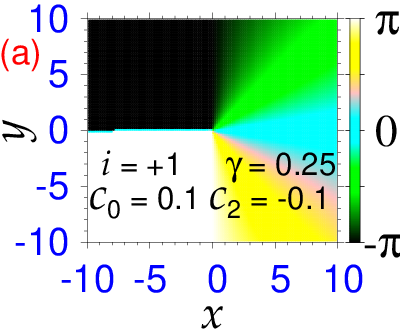}
\includegraphics[trim = 0mm 0mm 0cm 0mm, clip,height=3.cm,width= 3.1cm,clip]{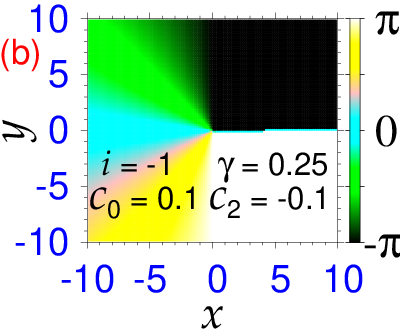}
\includegraphics[trim = 0mm 0mm 0cm 0mm, clip,height=3.cm,width= 3.cm,clip]{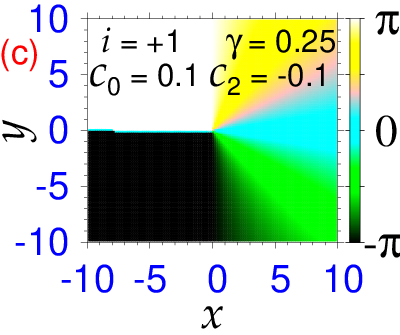}
\includegraphics[trim = 0mm 0mm 0cm 0mm, clip,height=3.cm,width= 3.1cm,clip]{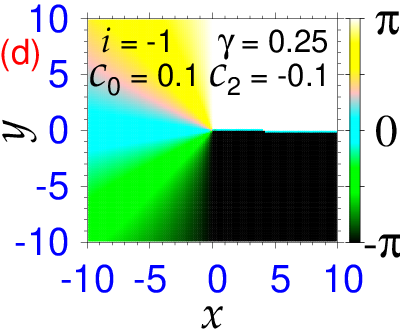}
\includegraphics[trim = 0mm 0mm 0cm 0mm, clip,height=3.cm,width= 3.cm,clip]{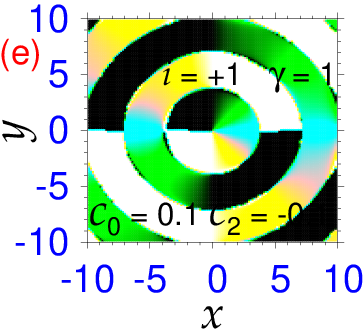}
\includegraphics[trim = 0mm 0mm 0cm 0mm, clip,height=3.cm,width= 3.1cm,clip]{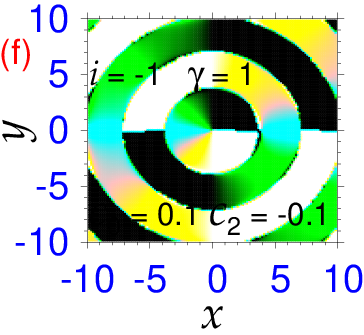}

\caption{(Color online)   Contour plot of the phase of the wave function of the symbiotic soliton shown in Figs. \ref{fig3}(a)-(b) for Rashba SO coupling of components (a) $i =+1$ and (b) $i =-1$.   The same of the symbiotic soliton shown in Figs. \ref{fig3}(a)-(b) for Dresselhaus SO coupling of components (c)  $i =+1$ and (d) $i =-1$.   Contour plot of the phase of the wave function of the symbiotic soliton shown in Figs. \ref{fig3}(c)-(d) for Rashba SO coupling of components (e) $i =+1$ and (f) $i =-1$.  }
\label{fig4} \end{center}
\end{figure}

\begin{figure}[!t]
\begin{center}
\includegraphics[trim = 0mm 0mm 0cm 0mm, clip,width=.47\linewidth,clip]{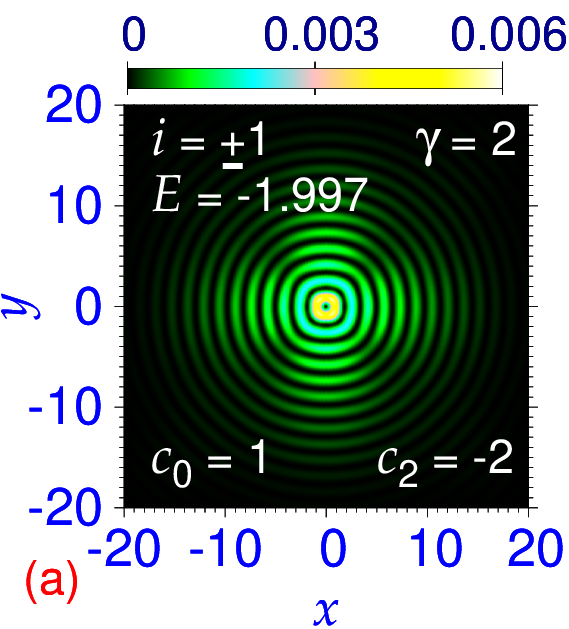}
\includegraphics[trim = 0mm 0mm 0cm 0mm, clip,width=.47\linewidth,clip]{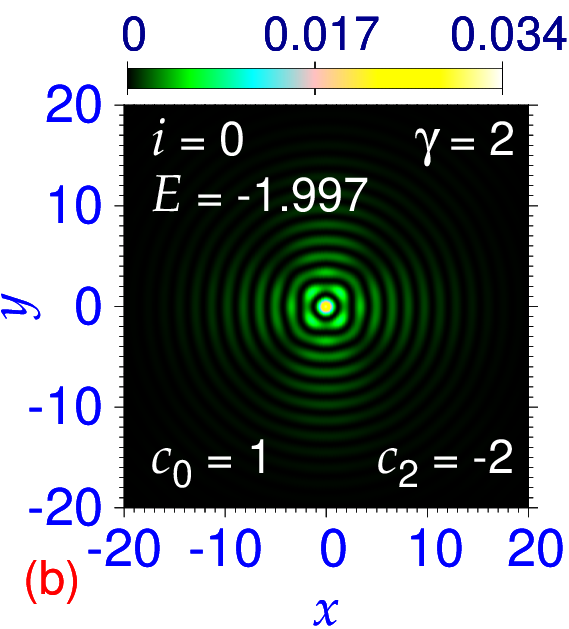}
\includegraphics[trim = 0mm 0mm 0cm 0mm, clip,width=.325\linewidth,clip]{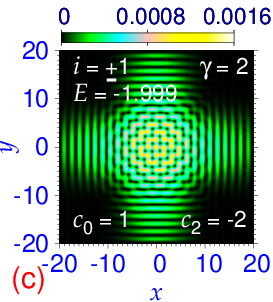}
\includegraphics[trim = 0mm 0mm 0cm 0mm, clip,width=.325\linewidth,clip]{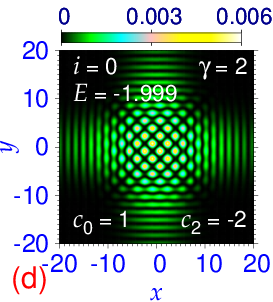}
\includegraphics[trim = 0mm 0mm 0cm 0mm, clip,width=.325\linewidth,clip]{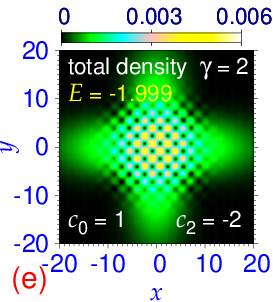} 
\includegraphics[trim = 0mm 0mm 0cm 0mm, clip,width=.325\linewidth,clip]{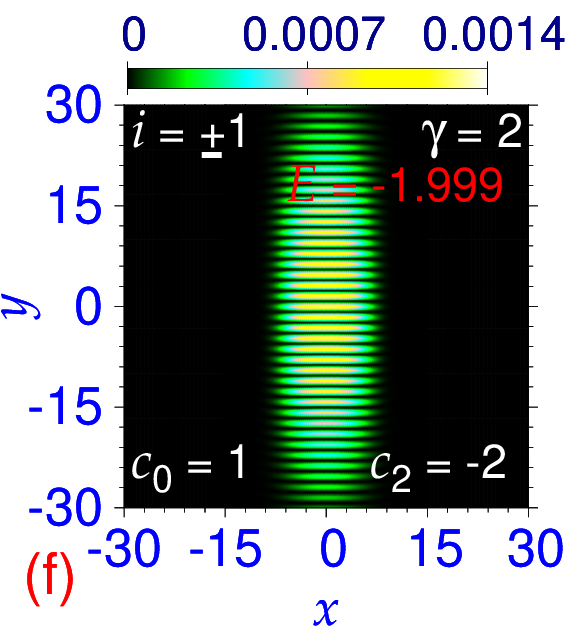}
\includegraphics[trim = 0mm 0mm 0cm 0mm, clip,width=.325\linewidth,clip]{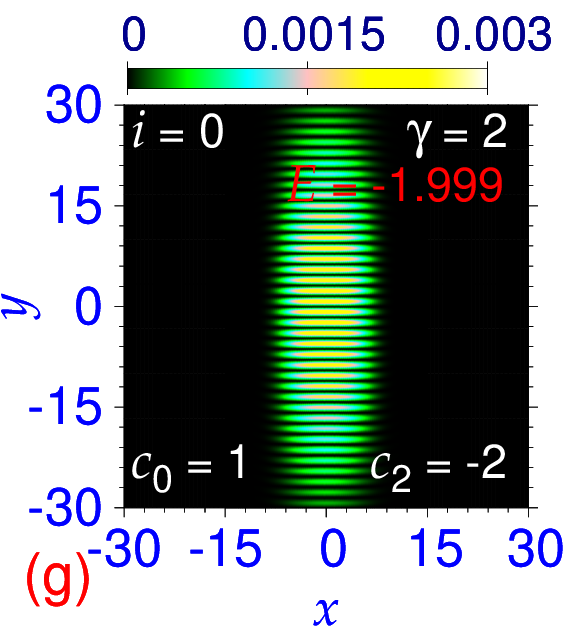}
\includegraphics[trim = 0mm 0mm 0cm 0mm, clip,width=.325\linewidth,clip]{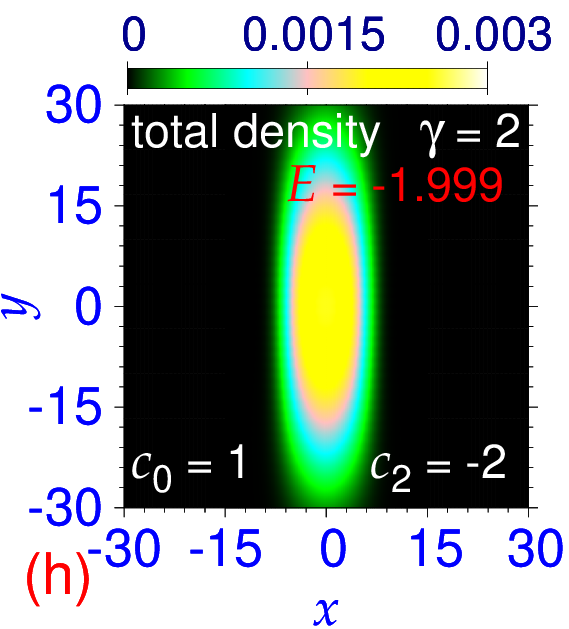} 

\caption{(Color online)   Contour plot of density $n_i$ of a quasi-2D  multi-ring  spin-1 Rashba
or Dresselhaus SO-coupled symbiotic BEC soliton of components (a)
$i = \pm 1$ and (b) $i = 0$; the same of
a superlattice soliton of components (c) $i = \pm 1$, (d) $i = 0,$
and (e) the total density; the same of
a stripe soliton of components (f) $i = \pm 1$, (g) $i = 0,$
and (h) the total density.  In all cases $c_0=1, c_2  =-2, \gamma =2$.}
\label{fig5} \end{center}
\end{figure}

As $\gamma$ is increased, a multi-ring structure appears in the densities of a  $(\mp 1,0,\pm 1)$-type  quasi-2D symbiotic SO-coupled  spin-1  spinor soliton
as illustrated in Fig. \ref{fig3} for $\gamma=1$ through a contour plot of density of components 
(c) $i =\pm 1$, (d) $i =0$ for    parameters
$c_0=1, c_2=-2$.  
 In addition to the  multi-ring symbiotic soliton, we find a new type of quasi-degenerate superlattice symbiotic soliton  as shown in Fig. \ref{fig3} for $\gamma =1$,
 where we display a contour plot 
of densities of components (e) $i =\pm 1$, (f) $i =0$ and (g) total density.  In this superlattice symbiotic soliton, a square-lattice structure  appears not only  in the component densities, viz. Figs. \ref{fig3}(e)-(f), but also in the total density,  viz. Fig. \ref{fig3}(g), thus illustrating the super-solid nature of this symbiotic soliton. In case of $\gamma=1$, the numerical energy of the superlattice soliton ($E=-0.499$) is approximately equal to the same of the multi-ring  ($E=-0.498$) soliton  and these two states are quasi degenerate.    In the numerical calculation by imaginary-time propagation the square-lattice structure was imprinted in the respective initial 
wave-function components.

\begin{figure}[!t]
\begin{center}

\includegraphics[trim = 0mm 0mm 0cm 0mm, clip,width=.325\linewidth,clip]{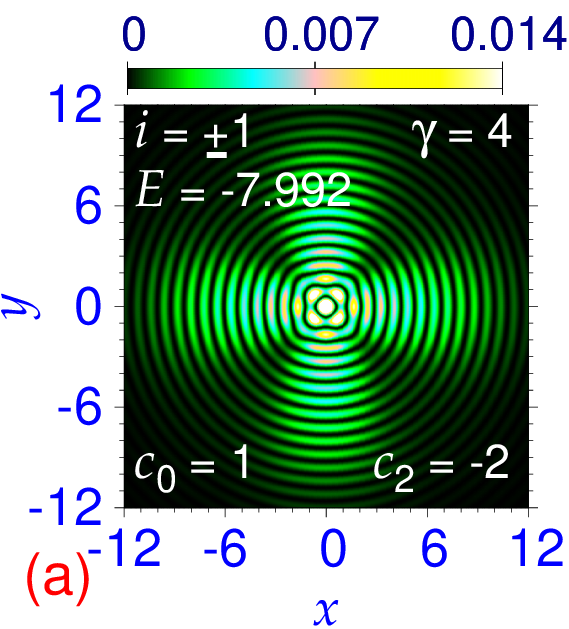}
\includegraphics[trim = 0mm 0mm 0cm 0mm, clip,width=.325\linewidth,clip]{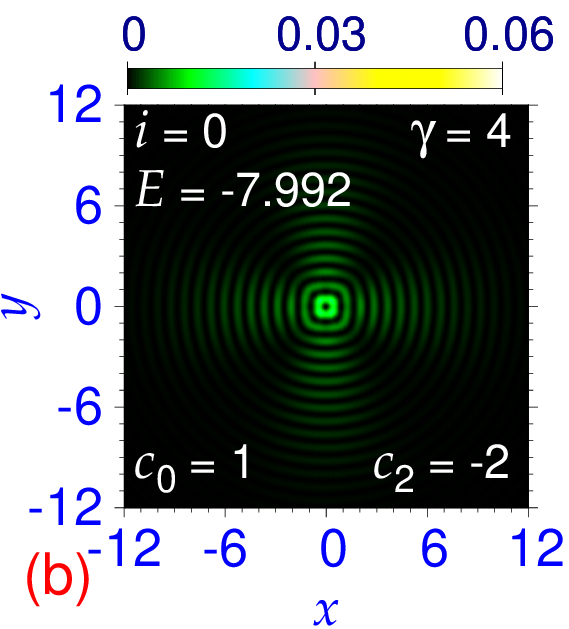} 
\includegraphics[trim = 0mm 0mm 0cm 0mm, clip,width=.325\linewidth,clip]{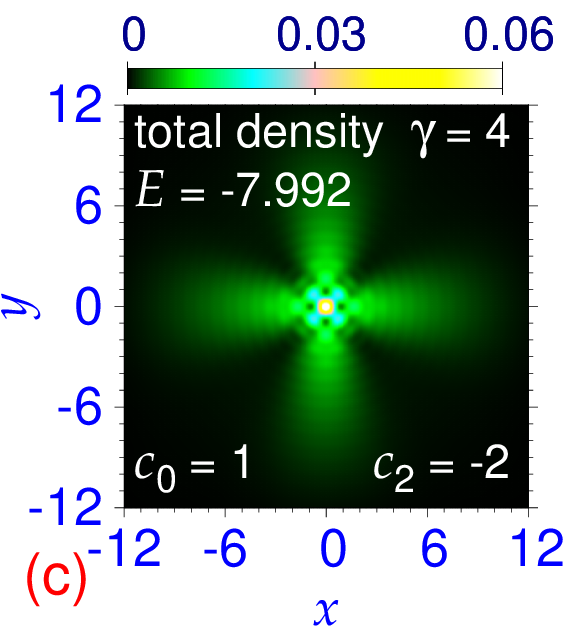}
\includegraphics[trim = 0mm 0mm 0cm 0mm, clip,width=.325\linewidth,clip]{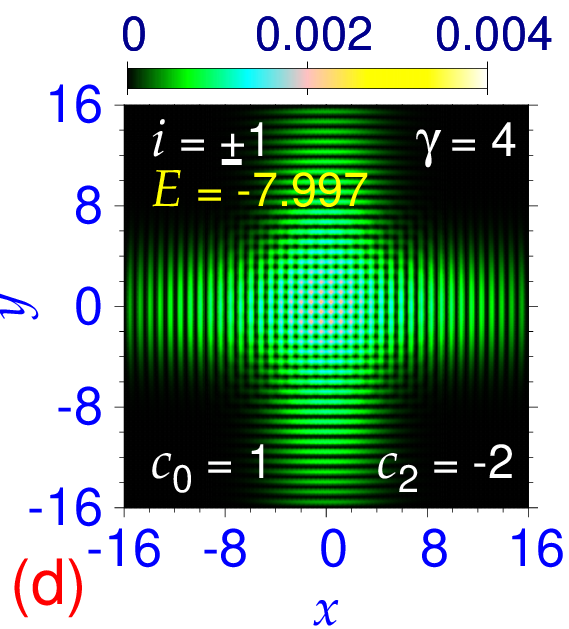} 
\includegraphics[trim = 0mm 0mm 0cm 0mm, clip,width=.325\linewidth,clip]{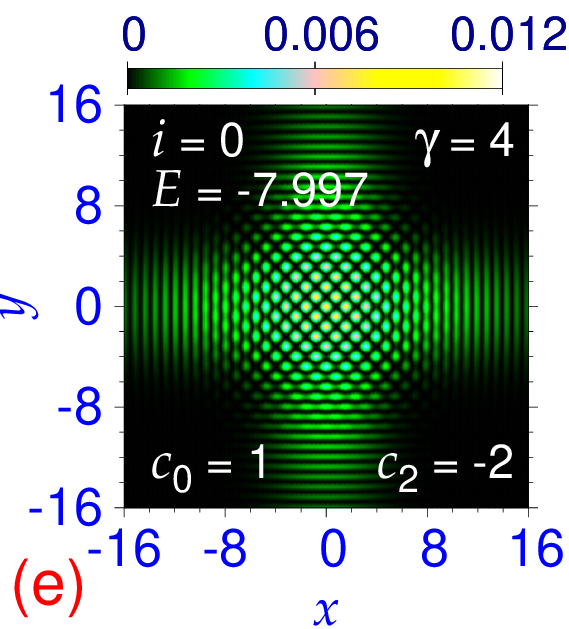} 
\includegraphics[trim = 0mm 0mm 0cm 0mm, clip,width=.55\linewidth,clip]{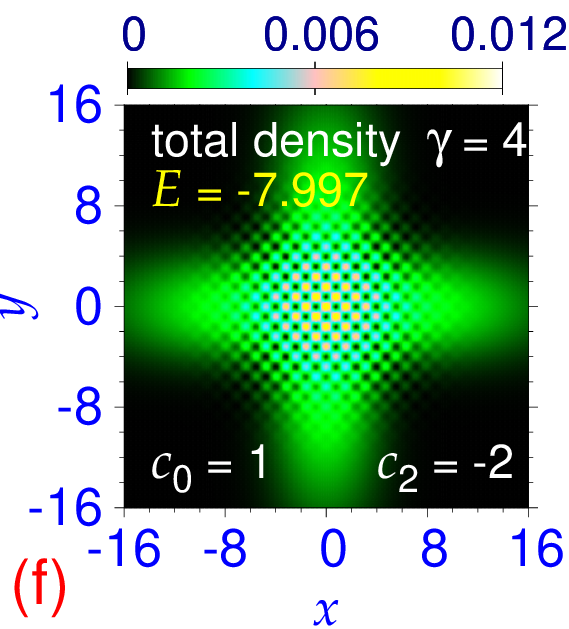}

\includegraphics[trim = 0mm 0mm 0cm 0mm, clip,width=.425\linewidth,clip]{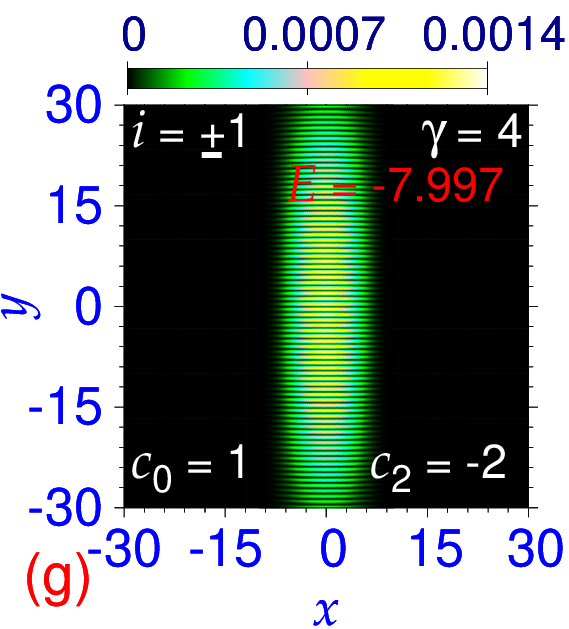} 
\includegraphics[trim = 0mm 0mm 0cm 0mm, clip,width=.425\linewidth,clip]{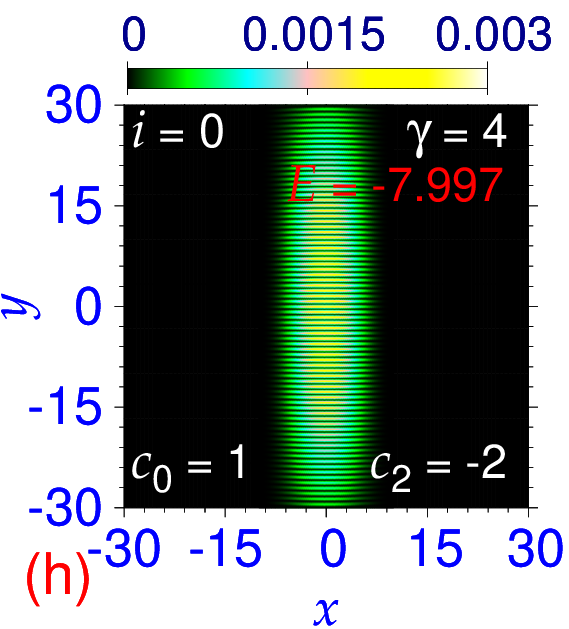}

\caption{(Color online)   Contour plot of density $n_i$ of a quasi-2D multi-ring spin-1 Rashba or Dresselhaus SO-coupled symbiotic BEC soliton of components (a) $i = \pm 1$, (b) $i = 0$ and
(c) total density; the same of a superlattice soliton of components (d) $i = \pm 1$, (e) $i = 0$, and (f) the total density; the
same of a stripe soliton of components (g) $i = \pm 1$ and (h)
$i = 0$.  In all cases $c_0=1, c_2  =-2, \gamma =4$.}
\label{fig6} \end{center}
\end{figure}

The presence of the vortices in the component densities of Figs. \ref{fig3}(a)-(d)  can be confirmed from a
study of the phase of the respective wave functions.  In  Fig. \ref{fig4} we show a contour plot of the phase of the wave-function components (a) $i =+1$  and (b) $i =-1$   of the soliton of Figs. \ref{fig3}(a)-(b) for  $\gamma =0.25$ for  Rashba SO coupling. The same for Dresselhaus SO coupling is shown in  Figs. \ref{fig4}
(c) $i =+1$  and (d) $i =-1$.  In Fig. \ref{fig4}(a) there is a phase drop of $- 2\pi$  upon  a complete rotation around the center corresponding to an antivortex of unit angular momentum at the center of   component $i =+ 1$ 
for Rashba SO coupling. In Fig. \ref{fig4}(b) the phase drop is $+2\pi$  corresponding to a vortex of unit angular momentum at the center of component $i =-1$.  The phase drops   upon  a complete rotation around the center
for Dresselhaus SO coupling in Figs. \ref{fig4}(c)-(d) are of opposite sign compared to those in Figs. \ref{fig4}(a)-(b) for Rashba SO coupling, respectively,  corresponding to a vortex (antivortex) of unit angular momentum in component $i =+1$  ($i =-1$).   
A contour plot of the phase of the wave function of components  $i =+1$  and  $i =-1$   of the soliton of Figs. \ref{fig3}(c)-(d)  for  Rashba SO coupling  with $\gamma=1$  is displayed in Figs. \ref{fig4}(e)-(f).
From the plots in  Figs. \ref{fig4}(e)  and (f) we confirm a phase drop of $-2\pi$ and $2\pi$, respectively, at the center corresponding to an antivortex (vortex)  in component $i =+1$  ($i=-1$) for Rashba coupling.  The phase drop for Dresselhaus coupling for $\gamma=1$ is of opposite sign, compared to Rashba coupling, 
corresponding to a vortex (antivortex) at the center of
component $i =+1$ ($i=-1$)   (not shown in this paper).

In Fig. \ref{fig5}  we display the density of components  (a) $i =\pm 1,$ and (b) $i=0$ of a $(\mp 1,0,\pm 1)$-type quasi-2D symbiotic multi-ring soliton
for  $c_0=1,c_2=-2$ and $\gamma=2$.  The circular symmetry of this multi-ring soliton is partially broken 
with the increase of $\gamma$.  In Fig. \ref{fig5} the density of components (c) $i =\pm 1,$ (d) $i=0$ and (e) total density of the superlattice soliton  for the same parameters is presented.  { In this case we also display in Fig. \ref{fig5} the contour plot of the density of the  stripe soliton     of components (f) $i=\pm 1$, (g) $i=0$, and (h) the total density. The total density in this case does not have any spatially-periodic modulation. 
 The energy of the multi-ring, superlattice and stripe solitons are  $E=-1.997, -1.999$ and $E=-1.999$ ($\approx -\gamma^2/2=-2$), respectively, so that the superlattice and the stripe solitons are quasi degenerate  ground states and the multi-ring soliton is a metastable  excited state.

In Fig. 6 we present the density of components (a) $i = \pm 1$, (b) $i = 0$ and (c) total density of the symbiotic multi-ring soliton for $ c_0 = 1, c_2 = -2$ and $\gamma  = 4$.
The circular symmetry of this multi-ring soliton is broken with the increase of $\gamma $. The density of the circularly symmetric
$\gamma  = 4$ state of Figs. 6(a)-(c) has acquired a  four-wing shape. 
In Fig. \ref{fig6} the density of components (d) $i =\pm 1$,  (e) $i=0$ and  (f) total density of the symbiotic superlattice soliton
for  $c_0=1,c_2=-2$ and $\gamma=4$ is displayed.  
 In Fig. \ref{fig6} the density of components (g) $i =\pm 1,$ and (h) $i=0$  of the   symbiotic stripe  soliton  for the same parameters  is presented.  The total density (not shown here) of the stripe soliton does not have any spatially-periodic modulation. The energies of the multi-ring, superlattice and stripe solitons are $E = -7.992, E = -7.997, E = -7.997$   ($\approx -\gamma^2/2=-8$), respectively,  so that the superlattice and the stripe solitons continue as quasi-degenerate ground states for large
SO coupling. 
The results of density and energy  presented in Fig. \ref{fig3}, \ref{fig5} and \ref{fig6} remain the same for both Rashba and Dresselhaus SO couplings, although the complex wave functions for these  SO couplings are different.

 \begin{figure}[!t]
\begin{center}
\includegraphics[trim = 0mm 0mm 0cm 0mm, clip,width=\linewidth,clip]{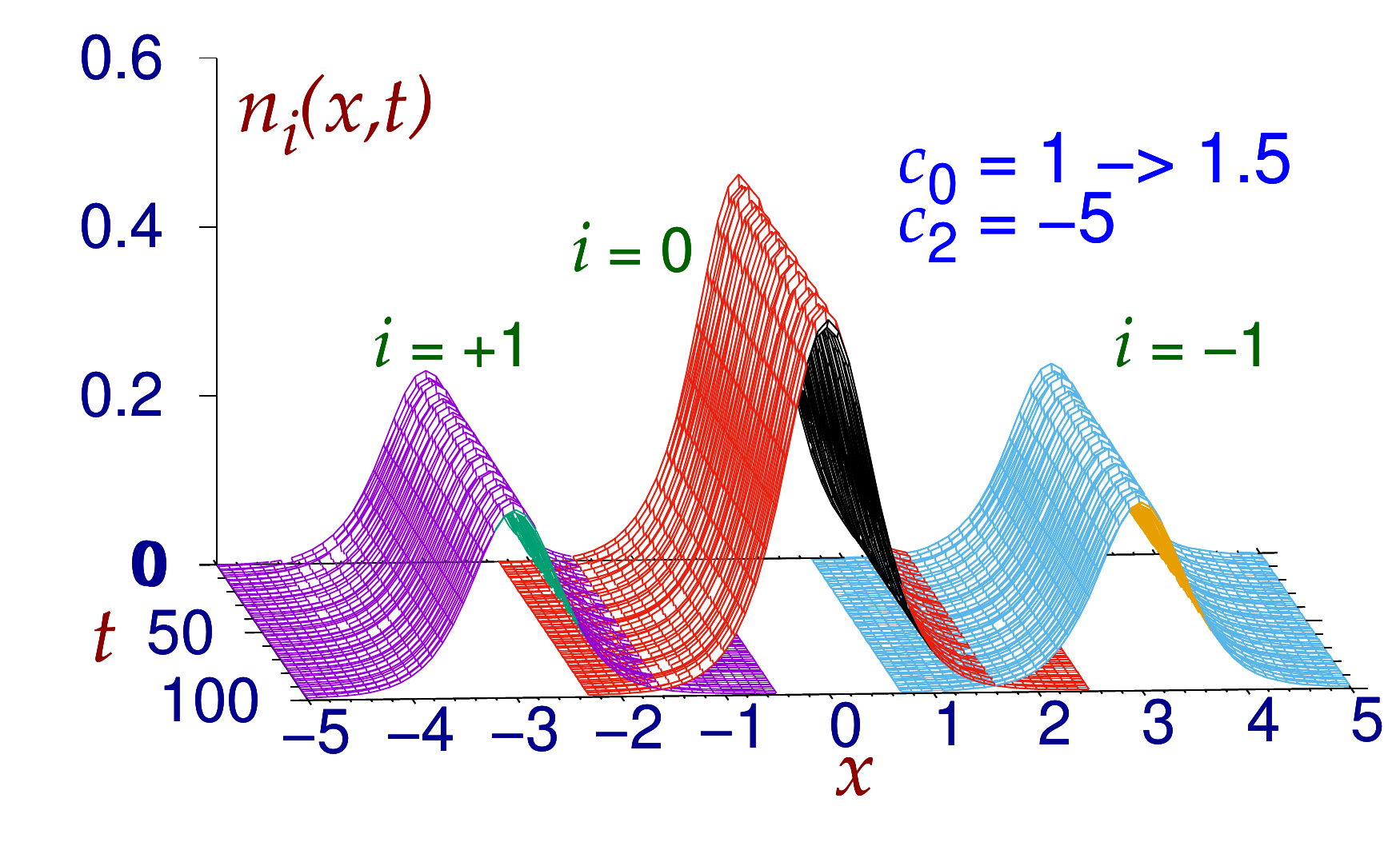}

\caption{(Color online) Density profile of the quasi-1D symbiotic spin-1 spinor  soliton $n_i(x,t)$
of Fig. \ref{fig2}(b) versus $x$ and $t$  during real-time propagation demonstrating its dynamical stability.
For a better view of the individual components, the component densities are plotted after a spatial displacement among these. 
The initial wave function was obtained by imaginary-time simulation with parameters 
$c_0=1, c_2=-5,$ and the real-time propagation using that initial wave function was executed after 
changing $c_0$ from 1 to 1.5 at $t=0$.
 }

\label{fig7} \end{center}
\end{figure}

 { In fact, with the increase of SO coupling,  the  $(\mp 1, 0, \pm 1)$-type  multi-ring soliton for $\gamma =4$, viz. Figs.  \ref{fig6}(a)-(c),  has turned to   an excited metastable state form a stable ground state for $\gamma =0.25$, viz. Figs. \ref{fig3}(a)-(b).  In imaginary-time propagation, after a very large number of time iterations, the multi-ring soliton becomes the ground superlattice soliton of Figs. \ref{fig6}(d)-(f). The total density of the metastable multi-ring soliton of  Fig. \ref{fig6}(c), after a reasonably large number of iterations,  has already acquired a square-lattice structure in the central region resembling a superlattice soliton,  indicating the beginning of a spontaneous transition to a superlattice state.  Hence, as the SO coupling increases, the $(\mp 1,0,\pm 1)$-type 
solitons spontaneously  break the rotational symmetry and become superlattice solitons.  It seems natural that the superlattice solitons can be  generated in imaginary-time propagation  from a square-lattice imprinted initial state (\ref{spf4}). But the fact that the use of a $(\mp 1, 0,\pm 1)$-type initial state for large SO coupling also leads to a superlattice soliton after a very large number of time iterations establishes the superlattice soliton as the robust ground state for large SO coupling,
 independent of finite system size and boundary condition, with the stripe soliton, with completely different symmetry, appearing as a quasi-degenerate state.  }

\begin{figure}[!t]
\begin{center}
\includegraphics[trim = 0mm 0mm 0cm 0mm, clip,width=\linewidth,clip]{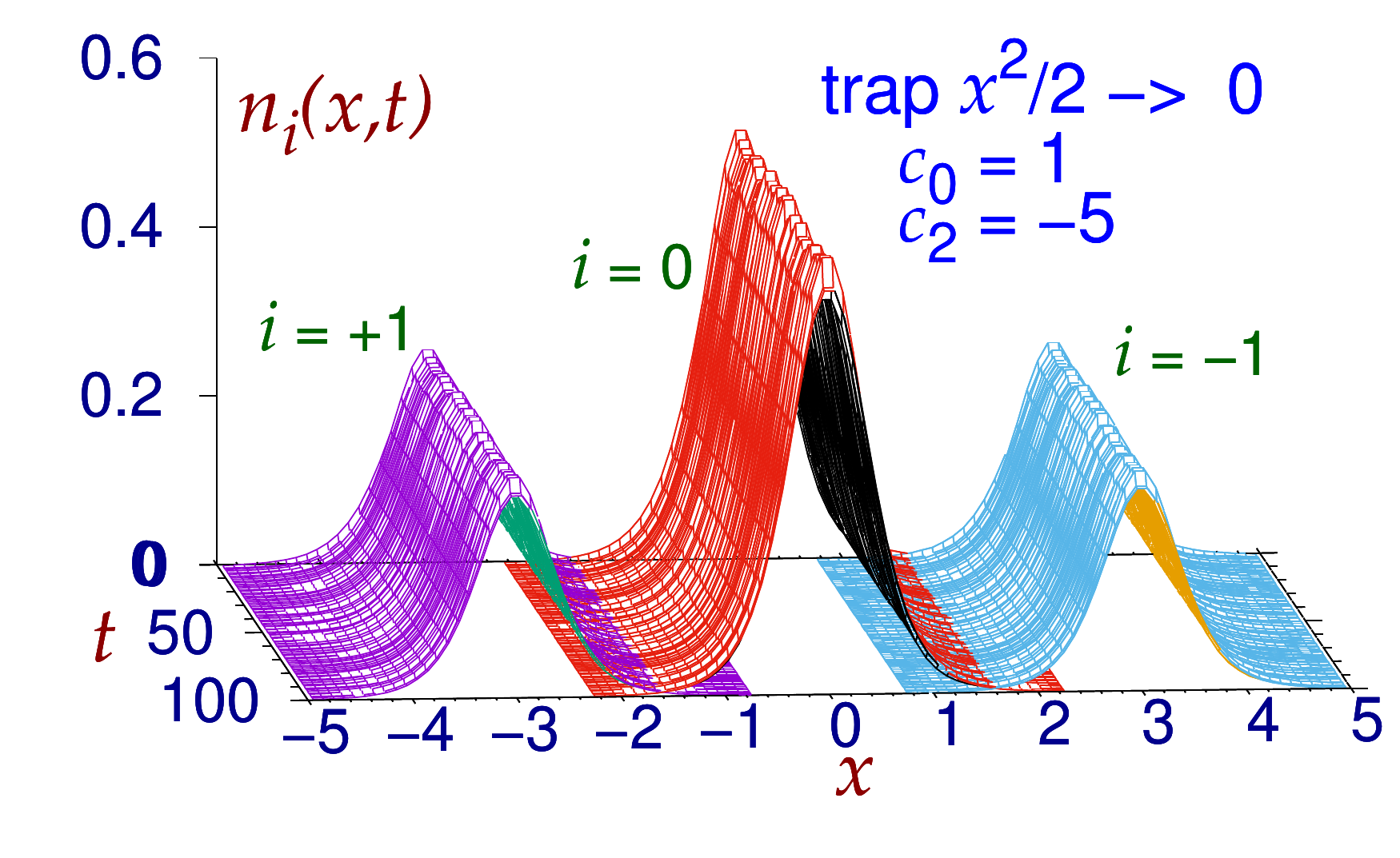}

\caption{(Color online) Density profile of a quasi-1D trapped symbiotic BEC  $n_i(x,t)$ 
with $c_0=1, c_2=-5$, viz. Fig. \ref{fig2}(b),  versus $x$ and $t$
during real-time propagation  after suddenly removing the trap  $V(x)=x^2 /2$ at $t=0$. The initial wave function in real-time propagation is the converged imaginary-time wave  function.
For a better view of the individual components, the component densities are plotted after a spatial displacement among these. 
 }

\label{fig8} \end{center}
\end{figure}
\subsection{Dynamical stability}

\subsubsection{Quasi-1D symbiotic spin-1  soliton}

\label{3C}

To demonstrate that the symbiotic vector soliton is dynamically stable we subject the ground-state vector soliton profile, 
obtained by imaginary-time simulation, to real-time propagation for a long time after giving a perturbation by 
changing the { interaction strength $c_0$ slightly at time $t=0$. The profile of the vector soliton is  sensitive to $c_0$. 
 }  For this purpose, we consider the quasi-1D symbiotic  
 vector soliton displayed in Fig. \ref{fig2}(b)
obtained with parameters $c_0=1, c_2=-5$. The real-time propagation during 100 time units 
for this soliton was 
executed upon  {changing the interaction strength $c_0 $ from 1 to 1.5 at time $t=0$.}     In Fig. \ref{fig7}  we exhibit the density profile of the three 
components of the vector soliton during real-time propagation. For a better view, we have displaced the density profile 
of components $i =+1$ and $i =-1$ to $x=-3$ and $x=+3$, respectively, leaving the  $i =0$ component at $x=0$. The 
long-time stable propagation of the components of the vector soliton establishes its dynamical stability.

\begin{figure}[!t]
\begin{center}
\includegraphics[width=.44\linewidth]{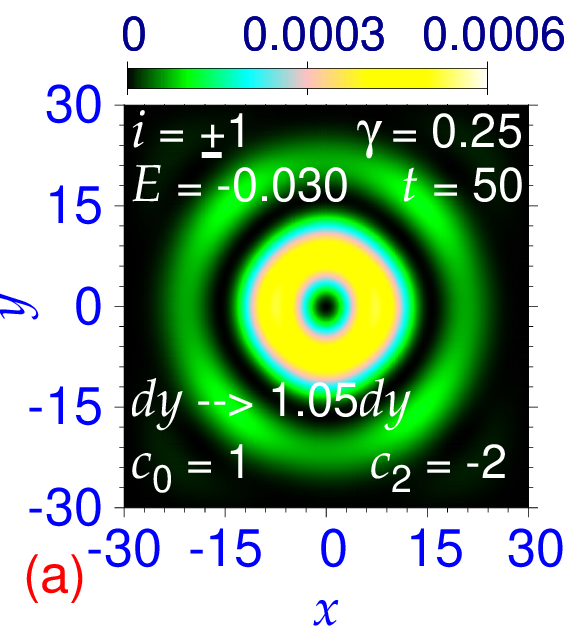} 
\includegraphics[width=.44\linewidth]{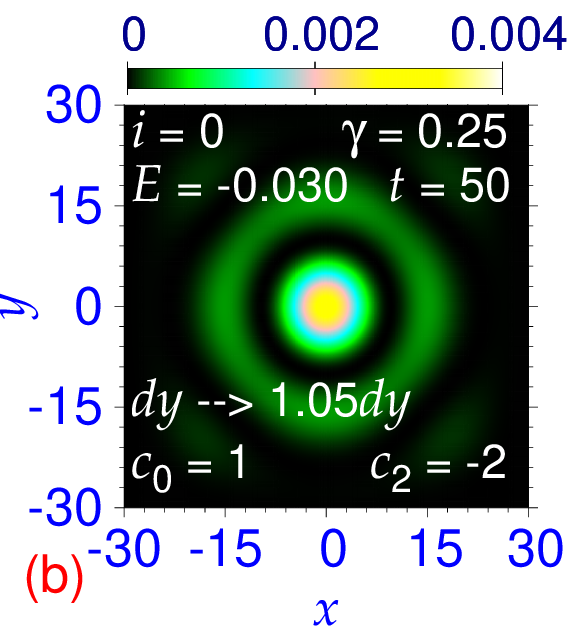}
\includegraphics[width=.325\linewidth]{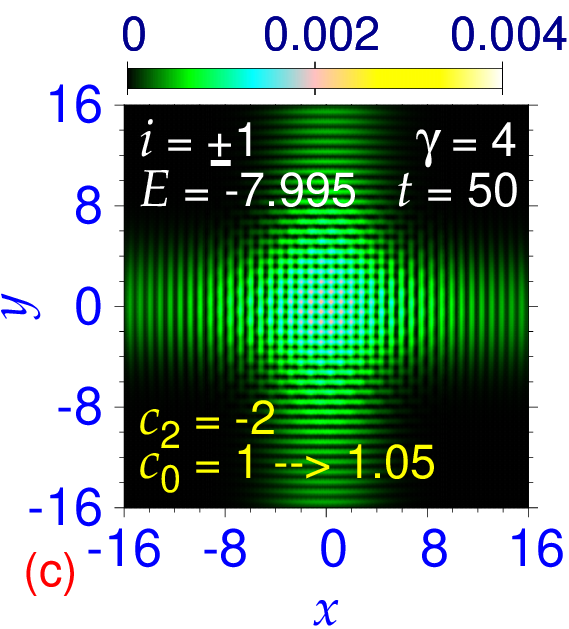}
\includegraphics[width=.325\linewidth]{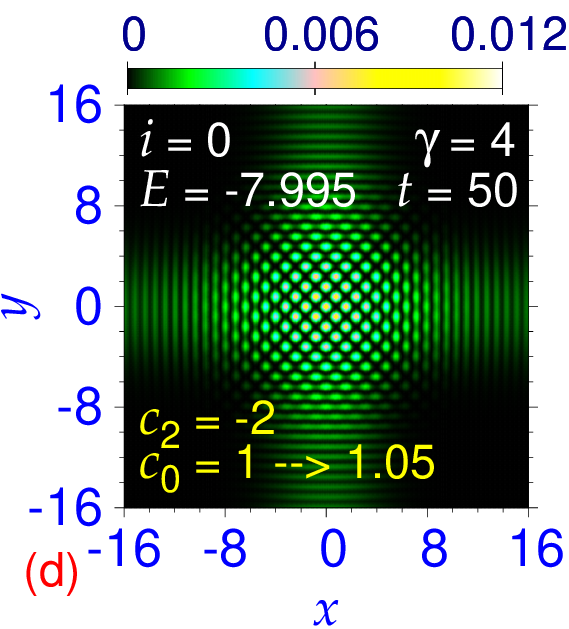}
\includegraphics[width=.325\linewidth]{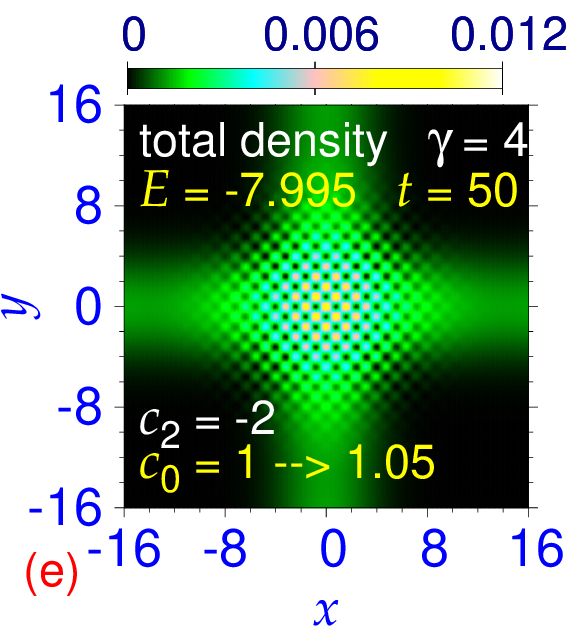}

\caption{(Color online)   Contour plot of density of the  quasi-2D symbiotic SO-coupled 
$(-1,0,+1)$-type  soliton with Rashba coupling, displayed in Figs. \ref{fig3}(a)-(b),
of components (a) $i = \pm 1$ and  (b) $i = 0$  after real-time propagation at time $t=50$ 
with a deformed wave function (see text for details); the  same of the  
superlattice symbiotic  BEC soliton, of
Figs. \ref{fig6}(d)-(f), of components (c) $i = \pm 1$, (d) $i = 0$ and (e)
the total density, after real-time propagation at time $t = 50$
upon a change of the nonlinearity coefficient $c_0$ from 1 to 1.05 at $t = 0$. }
\label{fig9} \end{center}
\end{figure}

Now we demonstrate  the possibility of the creation of the quasi-1D symbiotic soliton of Fig. \ref{fig2}(b) in a laboratory.  For this purpose we consider a quasi-1D symbiotic spin-1 BEC in a trap $V(x) = x^2/2$. First we consider the formation of this trapped BEC with $c_0=1$ and $c_2=-5$  by imaginary-time propagation. The density of this trapped BEC is plotted in Fig.  \ref{fig2}(b).  The resultant density is quite close to that of the untrapped soliton shown in this plot.
After creation,  this trapped BEC is subjected to real-time propagation upon suddenly removing the confining trap at time $t=0$.  The steady real-time propagation of the resultant system for 100 units of time  shown in Fig. \ref{fig8}  demonstrates both the dynamical stability of the symbiotic spin-1 soliton and 
 the plausibility of the creation of  the same  in a laboratory.

\subsubsection{Quasi-2D symbiotic SO-coupled spin-1 soliton}

\label{3D}

To demonstrate the dynamical stability of a quasi-2D symbiotic SO-coupled spin-1 spinor soliton,
we consider the   $(-1,0,+1)$-type   soliton of  { Figs. \ref{fig3}(a)-(b) and  the superlattice soliton  of
Figs. \ref{fig6}(d)-(f), using Rashba coupling. Of these, the  $(-1,0,+1)$-type
soliton of Figs. \ref{fig3}(a)-(b) hosting vortices is 
of special concern as vortices may exhibit transverse instability arising from a deformation of the wave function. To check the stability of this $(-1,0,+1)$-type  
soliton, we deform the soliton wave function slightly, so that the isodensity lines of  Figs. \ref{fig3}(a)-(b) are distorted to have an approximate elliptic shape, by increasing the space discretization step in the $y$ direction $dy$ by a factor of 1.05. Then we subject the imaginary-time
wave function in this deformed space  to real-time propagation during 50 units
of time. The resultant
density of the  $(-1,0,+1)$-type  SO-coupled quasi-2D     soliton at  $t=50$, displayed in Fig. \ref{fig9} for
components (a) $i = \pm 1$ and  (b) $i = 0$,  does not exhibit any transverse instability.  Starting from a radially-deformed shape the  $(-1,0,+1)$-type soliton has settled to its original circular-symmetric shape.}

To check the stability of the superlattice soliton of Figs. \ref{fig6}(d)-(f), we subject the imaginary-time wave function  to real-time propagation during 50 units
of time
 after changing the intraspecies   nonlinearity coefficient $c_0$  from 1 to 1.05 at time $t=0$. The resultant
density of the superlattice  soliton is displayed in Fig. \ref{fig9} for
components (c) $i = \pm 1$, (d) $i = 0$, and (e) total density.  In both cases, although the root-mean-square sizes and
energy were oscillating a little  during real-time propagation, the prominent pattern in
density survived at $t = 50$ as  can be seen in Fig. \ref{fig9}. If the solitons were not
dynamically stable, the vortex and superlattice structure of the wave functions  would have been  destroyed after real-time propagation.

\begin{figure}[!t]
\begin{center}
\includegraphics[width=.44\linewidth]{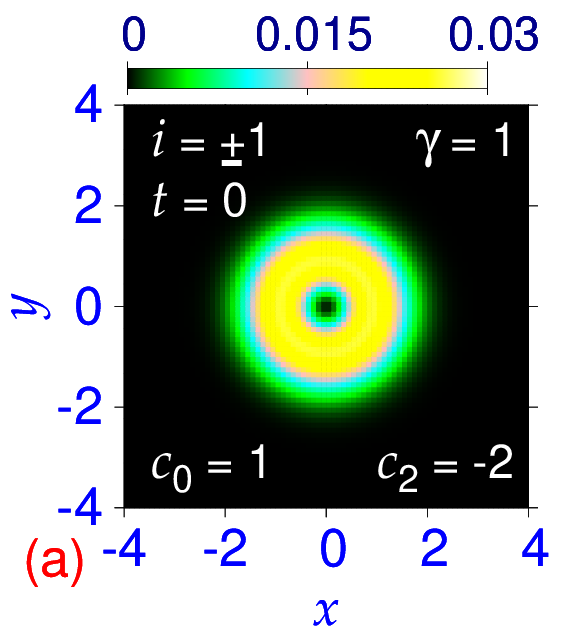} 
\includegraphics[width=.44\linewidth]{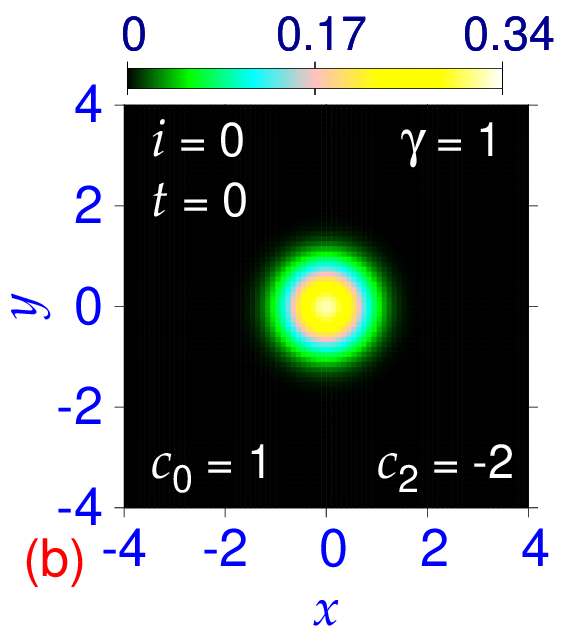}
\includegraphics[width=.44\linewidth]{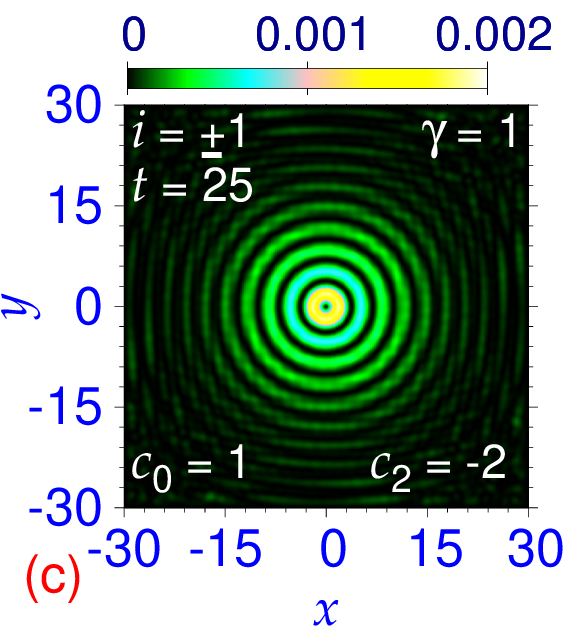}
\includegraphics[width=.44\linewidth]{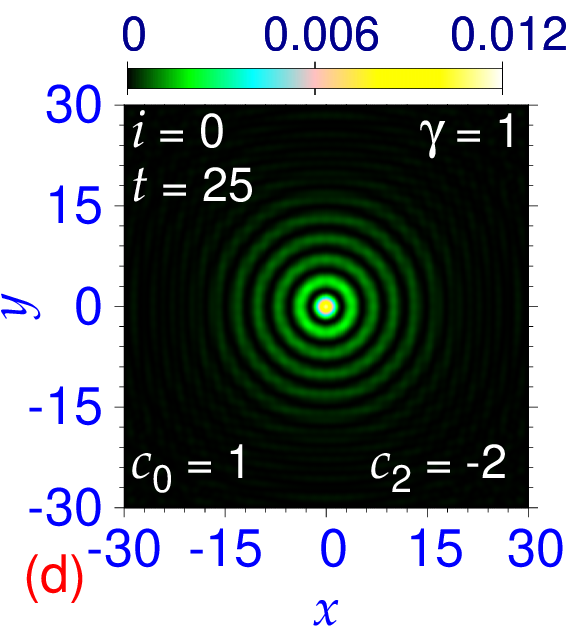}

\caption{(Color online) Contour plot of density of a  quasi-2D symbiotic SO-coupled 
$(-1,0,+1)$-type  trapped BEC  with Rashba coupling, 
of components (a) $i = \pm 1$ and  (b) $i = 0$ at $t=0$; the  same of the generated  $(-1,0,+1)$-type multi-ring   soliton after real-time propagation at time $t=25$
  of components (c) $i = \pm 1$ and  (d) $i = 0$,
upon a sudden removal of the trap at  $t = 0$.  The parameters of simulation: 
$c_0=1, c_2=-2, \gamma=1$  }
\label{fig10} \end{center}
\end{figure}

Finally, we consider  the possibility of the creation of a quasi-2D $(\mp 1,0,\pm 1)$-type  multi-ring
symbiotic soliton of Figs. \ref{fig3}(c)-(d) in a laboratory.  For this purpose we prepare a quasi-2D symbiotic spin-1 BEC, in a trap $V({\bf r}) = r^2/2$,  with  parameters $c_0=1$ and $c_2=-2$ and Rashba SO-coupling strength $\gamma =1$,  by imaginary-time propagation. The contour plot of density of this trapped BEC is plotted, in Fig.  \ref{fig10},  of components (a) $i=\pm 1$ and (b) $i=0$.   
After creation,  this trapped BEC is subjected to real-time propagation upon a sudden removal of  the confining trap at time $t=0$.  A snapshot of the contour density of the  generated soliton at time $t=25$  is displayed, in Fig. \ref{fig10},  of components (c) $i=\pm 1$ and (d) $i=0$. During real-time propagation the size of the BEC has increased by a factor of 10, while the rings were generated and the BEC was being  transformed to a multi-ring soliton. 
 The generated soliton is quite similar to the same of  Figs. \ref{fig3}(c)-(d), which demonstrates the robustness and stability of this soliton.

\section{Summary}
\label{Sec-IV}

We studied the formation of  quasi-1D and quasi-2D symbiotic spin-1 spinor solitons in a self-repulsive spin-1 BEC  using a numerical solution 
and an analytic approximation of the underlying mean-field GP equation with two nonlinear parameters $c_0$ 
and $c_2$.  A symbiotic soliton with intraspecies repulsion and interspecies attraction should necessarily require $c_0>0$ and $c_2<0$.  Even then there is a self-attractive term proportional to $c_2 n_{\pm} \psi_{\pm}$ in components $i =\pm 1$, viz. Eqs. \ref{gps-1}  and \ref{gps-4}. This  inappropriate self-attractive  term can be 
avoided if we further impose the condition $n_{+1}=n_{-1}$, so that      Eqs. \ref{gps-1}  and \ref{gps-4}
reduce to    Eqs. \ref{gps-2}  and \ref{gps-6}, respectively. Consequently,  solutions of
Eqs. \ref{gps-2}  and \ref{gps-3}, and Eqs. \ref{gps-6}  and \ref{gps-5}, in quasi-1D and quasi-2D configurations, respectively, are appropriate for the study of symbiotic quasi-1D and quasi-2D spin-1 solitons
with the property  $n_{+1}=n_{-1}$. In the quasi-1D case the numerical result for density and energy  of the symbiotic spin-1 spinor  soliton are in good agreement with an analytic approximation.

In the quasi-2D case, in addition, a non-zero SO coupling is necessary for the formation of a symbiotic spin-1 BEC soliton. For a small strength $\gamma$ of SO coupling, the quasi-2D symbiotic SO-coupled spin-1 spinor soliton is   
of the $(\mp 1, 0, \pm 1)$ type with multi-ring structure hosting vortices of vorticity 
$\mp 1$ and $\pm 1$ in components $i =+1$ and $-1$, respectively, where the upper (lower) sign refer to Rashba (Dresselhaus) SO coupling. For large strength of SO coupling, in addition to the above multi-ring symbiotic solitons there appear quasi-degenerate stripe  solitons and superlattice solitons.  The superlattice solitons have  square-lattice structure in component and total densities, thus  sharing  properties of a supersolid. 
{ For larger SO coupling, the  $(\mp 1, 0, \pm 1)$-type  multi-ring soliton becomes an excited metastable state and  spontaneously breaks the  rotational symmetry to become a stable superlattice soliton.   The superlattice and the  stripe solitons continue as  quasi-degenerate ground  states. 
  An SO-coupled quasi-1D spin-1 BEC naturally leads to an 1D periodic pattern in density \cite{29}; hence an SO-coupled  quasi-2D BEC involving two perpendicular SO-coupling directions, as considered in  this paper, should naturally lead to a square-lattice 
structure in density in addition to the stripe pattern. We could not rule out the possibility of the appearance of a triangular-lattice or other structure, although we could not  find out any other stable periodic structure in this study. }
In all cases, the energy and  density of the quasi-2D symbiotic soliton are the same for both Rashba and Dresselhaus SO couplings although the two complex wave functions are different. The dynamical stability of the quasi-1D and quasi-2D symbiotic spin-1 spinor solitons is demonstrated numerically by real-time propagation over a long period of time. The plausibility of the formation of a  symbiotic spin-1 BEC soliton in a laboratory  by removing the trap of a spin-1 trapped BEC is also demonstrated.


\begin{acknowledgements}

This work is financed by the  CNPq (Brazil) grant 301324/2019-0, and by the
ICTP-SAIFR-FAPESP (Brazil) grant 2016/01343-7.

\end{acknowledgements}

\end{document}